\documentclass[12pt]{article}
\usepackage{epsfig}

\begin{document}

\begin{titlepage}
\begin{flushright}
gr-qc/0505115 \\ UFIFT-QG-05-02 \\ CRETE-05-11
\end{flushright}

\vspace{0.5cm}

\begin{center}
\bf{STOCHASTIC QUANTUM GRAVITATIONAL INFLATION}
\end{center}

\vspace{0.3cm}

\begin{center}
N. C. Tsamis$^{\dagger}$
\end{center}
\begin{center}
\it{Department of Physics, University of Crete \\
GR-710 03 Heraklion, HELLAS.}
\end{center}

\vspace{0.2cm}

\begin{center}
R. P. Woodard$^{\ddagger}$
\end{center}
\begin{center}
\it{Department of Physics, University of Florida \\
Gainesville, FL 32611, UNITED STATES.}
\end{center}

\vspace{0.3cm}

\begin{center}
ABSTRACT
\end{center}
\hspace{0.3cm}
During inflation explicit perturbative computations of quantum
field theories which contain massless, non-conformal fields
exhibit secular effects that grow as powers of the logarithm of
the inflationary scale factor. Starobinski\u{\i}'s technique of
stochastic inflation not only reproduces the leading infrared
logarithms at each order in perturbation theory, it can sometimes
be summed to reveal what happens when inflation has proceeded so
long that the large logarithms overwhelm even very small coupling
constants. It is thus a cosmological analogue of what the
renormalization group does for the ultraviolet logarithms of
quantum field theory, and generalizing this technique to quantum
gravity is a problem of great importance. There are two significant
differences between gravity and the scalar models for which
stochastic formulations have so far been given: derivative
interactions and the presence of constrained fields. We use
explicit perturbative computations in two simple scalar models
to infer a set of rules for stochastically formulating theories
with these features.

\vspace{0.3cm}

\begin{flushleft}
\end{flushleft}

\vspace{0.1cm}

\begin{flushleft}
$^{\dagger}$ e-mail: tsamis@physics.uoc.gr \\
$^{\ddagger}$ e-mail: woodard@phys.ufl.edu
\end{flushleft}

\end{titlepage}

\section{Introduction}

During inflation, the energy-time uncertainty principle allows
any massless virtual particle that emerged from the vacuum to
persist forever. If classical conformal invariance is present,
the rate of emergence redshifts so that very few such virtual
particles are present. It is only for gravitons and minimally
coupled scalars that classical conformal invariance is broken
in such a way that inflation can give strong enhancements of
quantum effects. \\

$\bullet$ {\it de Sitter Inflation:} A locally de Sitter geometry
provides the simplest paradigm for inflation. To see why, consider
a general homogeneous, isotropic and spatially flat geometry:
\begin{equation}
ds^2 \; = \;
- \, dt^2 \, + \, a^2 (t) \, d{\vec x} \cdot d{\vec x}
\;\; . \label{ds2}
\end{equation}
Derivatives of the scale factor $a(t)$ give the Hubble parameter
$H(t)$ and the deceleration parameter $q(t)$:
\begin{equation}
H(t) \; \equiv \; \frac{\dot{a}}{a}
\qquad , \qquad
q(t) \; \equiv \; -\frac{a \ddot{a}}{\dot{a}^2}
\, = \, -1 - \frac{\dot{H}}{H^2}
\;\; . \label{Hq}
\end{equation}
The nonzero components of the Riemann tensor are:
\begin{equation}
R^0_{~i0j} = - q H^2 \, g_{ij}
\qquad , \qquad
R^i_{~jk\ell} = H^2 \Bigl(
\delta^i_{\, k} \; g_{j\ell} - \delta^i_{\, \ell} \; g_{jk} \Bigr)
\;\; . \label{Riemann}
\end{equation}
Inflation is defined as positive expansion ($H(t) > 0$) with
negative deceleration ($q(t) < 0$). On the other hand, stability
-- in the form of the weak energy condition -- implies $q(t) \geq -1$.
At the limit of $q = -1$ we see from (\ref{Riemann}) that the Riemann
tensor assumes the locally de Sitter form:
\begin{equation}
\lim_{q = -1}
R^{\rho}_{~\sigma \mu \nu} \, = \, H^2 \Bigl(
\delta^{\rho}_{\; \mu} \; g_{\sigma \nu} -
\delta^{\rho}_{\; \nu} \; g_{\sigma \mu} \Bigr)
\;\; . \label{dSRiemann}
\end{equation}
It follows from (\ref{Hq}) that the Hubble parameter is actually
constant, and that the zero of time can be chosen to make the scale
factor take the simple exponential form we shall henceforth assume:
\begin{equation}
{\it de \; Sitter \; Inflation}
\quad \Longrightarrow \quad
a(t) \, = \, e^{Ht}
\;\; . \label{dS}
\end{equation}

$\bullet$ {\it Particle Production:} The homogeneity of spacetime
expansion in (\ref{ds2}) does not change the fact that particles have
constant wave vectors $\vec{k}$, but it does alter their physical meaning.
In particular, the energy of a particle with mass $m$ and wave number
$k$ becomes time dependent:
\footnote{Of course ``energy'' is not a good quantum number in de Sitter
background. What we mean by ``$E(t, k)$'' is the function whose integral
times $(-i)$ determines the phase of plane wave mode functions. This 
statement is exact for $m = 0$ and true in the WKB limit for $m \neq 0$.} 
\begin{equation}
E(t,k) \, = \, \sqrt{m^2 + \frac{k^2}{a^2(t)}}
\qquad , \qquad
k \equiv \Vert \vec{k}\Vert
\;\; . \label{E}
\end{equation}
This results in an interesting change in the energy-time uncertainty
principle which restricts how long a virtual pair of such particles
with wave vectors $\pm \vec{k}$ can exist. If the pair was created
at time $t$, it can last a time $\Delta t$ given by the integral:
\begin{equation}
\int_t^{t+\Delta t} dt' \; E(t',k) \, \sim \, 1
\;\; . \label{dedt}
\end{equation}
Just as in flat space, particles with the smallest masses persist
longest. For the fully massless case, the integral is simple to evaluate,
\begin{equation}
\int_t^{t+\Delta t} dt' \;  E(t',k) \, \Biggl\vert_{m=0}
\, = \;
\Biggl[ \, 1 - e^{-H \Delta t} \, \Biggr] \,
\frac{k}{H a(t)}
\;\; . \label{m=0dedt}
\end{equation}
We, therefore, conclude that any massless virtual particle which
happens to emerge from the vacuum can persist forever provided:
\begin{equation}
{\it Unbounded \; Lifetime}
\quad \Longrightarrow \quad
k \leq H a(t)
\;\; . \label{inftlife}
\end{equation}

$\bullet$ {\it Conformal Invariance:} Most massless particles possess
conformal invariance. A simple change of variables defines a conformal
time $\eta$ in terms of which the invariant element (\ref{ds2}) is just
a conformal factor times that of flat space:
\begin{equation}
ds^2 \, = \,
- dt^2 \, + \, a^2 (t) \, d{\vec x} \cdot d{\vec x}
\, = \,
a^2(\eta) \, \Bigl(
- d\eta^2 \, + \, d{\vec x} \cdot d{\vec x} \, \Bigr)
\quad , \quad
d\eta \, \equiv \, \frac{dt}{a(t)}
\;\; . \label{eta}
\end{equation}
In the $(\eta,\vec{x})$ coordinates, conformally invariant theories
are locally identical to their flat space counterparts. The rate at
which virtual particles emerge from the vacuum per unit conformal
time must be the same constant -- call it $\Gamma$ -- as in flat
space. 
\footnote{For particles which do not possess conformal invariance,
the rates are generically different in de Sitter than in flat 
spacetime \cite{BV2,BV3}.}

Hence, the rate of emergence per unit physical time is:
\begin{equation}
\frac{dN}{dt} \, = \,
\frac{dN}{d\eta} \, \frac{d\eta}{dt}
\, = \, \frac{\Gamma}{a(t)}
\;\; . \label{rate}
\end{equation}
Consequently -- although any sufficiently long wavelength, massless
and conformally invariant particle emerging from the vacuum can
persist forever during inflation -- very few such particles will
actually emerge. \\

$\bullet$ {\it Quantum Enhancement:} Gravitons and minimally coupled
scalars are two kinds of massless particles which do not possess conformal
invariance. To see that -- unlike massless conformally invariant particles
-- the production of these two kinds of particles is not suppressed during
inflation, note that each polarization and wave number behaves like a
harmonic oscillator with time dependent mass and frequency:
\begin{equation}
L \, = \, \frac12 m \dot{q}^2 \, - \, \frac12 m \omega^2 q^2
\qquad , \qquad
m(t) = a^3(t)
\quad \& \quad
\omega(t) = \frac{k}{a(t)}
\;\; . \label{L1}
\end{equation}
The Heisenberg equation of motion can be exactly solved:
\begin{equation}
\ddot{q} \, + \, 3 H \, \dot{q} \, + \, \frac{k^2}{a^2} \, q \, = \, 0
\qquad \Longrightarrow \qquad
q(t) \, = \, u(t,k) \; \alpha \, + \, u^*(t,k) \; \alpha^{\dagger}
\;\; , \label{eom1}
\end{equation}
where the mode functions $u$ and the commutation relations obeyed
by the operators $\alpha$ and $\alpha^{\dagger}$ are given by:
\begin{equation}
u(t,k) \, = \, \frac{H}{\sqrt{2 k^3}} \Biggl[ \,
1 - \frac{i k}{H a(t)}\, \Biggr] \;
\exp \left( \frac{ik}{Ha(t)} \right)
\qquad , \qquad
[\, \alpha \, , \, \alpha^{\dagger} \, ] \, = \, 1
\;\; . \label{mode&cc}
\end{equation}
The co-moving energy operator for this system is:
\begin{equation}
E(t) \, = \, \frac12 m(t) \, \dot{q}^2(t) \, + \,
\frac12 m(t) \, \omega^2(t) \, q^2(t)
\;\, . \label{Eop}
\end{equation}
Owing to the time dependent mass and frequency, there are no stationary
states for this system. At any given time the minimum eigenstate of
$E(t)$ has energy $\frac12 \omega(t)$, but which state this is changes
for each value of time. The state $\vert \Omega \rangle$ which is
annihilated by $\alpha$ has minimum energy in the distant past. The
expectation value of the energy operator in its presence is,
\begin{equation}
\Bigl\langle \Omega \, \Bigl\vert \, E(t) \,
\Bigr\vert \, \Omega \Bigr\rangle  \, = \,
\frac12 a^3(t) \, \vert \dot{u}(t,k) \vert^2  \, + \,
\frac12 a(t) \, k^2 \, \vert u(t,k) \vert^2  \, = \,
\frac{k}{2a} \, + \, \frac{H^2 \, a}{4k}
\;\; . \label{EopVEV}
\end{equation}
If we think of each particle as having energy $k \, a^{-1}(t)$, it
follows that the number of particles $N$ with any polarization and
wave number $k$ grows as the square of the inflationary scale factor:
\begin{equation}
N(t,k) \, = \, \Biggl[ \,
\frac{H a(t)}{2 k} \, \Biggr]^2
\;\;. \label{Ngrowth}
\end{equation}

Quantum field theoretic effects are driven by essentially classical
physics operating in response to the source of virtual particles
implied by quantization. On the basis of (\ref{Ngrowth}), one might
expect inflation to dramatically enhance quantum effects from massless,
minimally coupled scalars and gravitons. This has been confirmed
explicitly and the oldest results are the cosmological perturbations
induced by scalar inflatons \cite{mukhanov} and by gravitons 
\cite{starobinsky1}. The more recent result which motivated the present 
analysis is that the gravitational back-reaction from the inflationary 
production of gravitons induces an ever greater slowing in the expansion 
rate \cite{nctrpw1,nctrpw2}. \\

$\bullet$ {\it Quantum Cosmology:} The Lagrangian is the two-parameter
effective gravitational theory:
\begin{equation}
{\cal L}_{\mbox{\tiny GR}} = {1 \over 16 \pi G}
\Bigl(- 2 \Lambda + R \; \Bigr) \; \sqrt{-g}
\quad , \quad
H^2 \equiv \frac13 \Lambda > 0
\;\; . \label{Lgr}
\end{equation}
What was actually computed \cite{nctrpw2} is the graviton one-point 
function, about a locally de Sitter background, in the presence of a 
state which is free Bunch-Davies vacuum at $t=0$. However, if the 
resulting distortion of the background geometry was viewed in terms 
of an effective energy density and pressure the perturbative infrared 
results would be:
\begin{eqnarray}
\rho(t) & = &
\frac{\Lambda}{8 \pi G} \; + \;
\frac{(\kappa H)^2 H^4}{2^6 \pi^4} \,
\Biggl\{ -\frac12 \ln^2 a \, + \, O\Bigl(\ln a\Bigr) \Biggr\}
\; + \; O(\kappa^4)
\;\; , \label{rhoGR} \\
p(t) & = &
-\frac{\Lambda}{8 \pi G} \; + \;
\frac{(\kappa H)^2 H^4}{2^6 \pi^4} \,
\Biggl\{ \, \frac12 \ln^2 a \, + \, O\Bigl(\ln a\Bigr) \Biggr\}
\; + \; O(\kappa^4)
\;\; , \label{pGR}
\end{eqnarray}
where $\kappa^2 \equiv 16 \pi G$ is the loop counting parameter of quantum
gravity. The one-loop effect from the kinematic energies of inflationary
gravitons is a constant \cite{ford1,finelli} that must be subsumed into
$\Lambda$. The next order effect is secular because the kinematic energies
interact with the total graviton field strength which grows as more and
more gravitons are produced. The induced energy density is negative because
the gravitational interaction is attractive.

One would like to improve the aforementioned computation on the level
of perturbation theory and beyond. \\

$\bullet$ {\it Invariant Regulation:} A perturbative improvement would be
to use $D$-dimensional Feynman rules and dimensional regularization to
obtain the fully renormalized answer and avoid complications that can arise
from non-invariant counterterms. Such a procedure has been used in the
simpler case of a massless, minimally coupled scalar field with a quartic
self-interaction of strength $\lambda$ in a non-dynamical, locally de
Sitter background:
\begin{equation}
{\cal L}_{\varphi} \; = \;
-\frac12 \, \sqrt{-g} \, g^{\mu\nu} \,
\partial_{\mu} \varphi \; \partial_{\nu} \varphi
\; - \; \frac{\lambda}{4!} \, \sqrt{-g} \; \varphi^4
\;\; . \label{Lphi4}
\end{equation}
This theory can be renormalized so that, when released in free
Bunch-Davies vacuum at $t=0$, the energy density and pressure are
\cite{rpw1,rpw2}:
\begin{equation}
\rho_{\mbox{\tiny{ren}}}(t) \, = \,
\frac{\Lambda}{8\pi G} \, + \,
\frac{\lambda H^4}{2^6\pi^4} \Bigg[ \,
\frac12 \ln^2 a \, + \,
\frac29 \, a^{-3} \, - \,
\frac12 \sum_{n=1}^{\infty} \frac{n+2}{(n+1)^2} \, a^{-n-1}
\Bigg] \; + \; O(\lambda^2)
\;\; . \label{rhophi4}
\end{equation}
\begin{equation}
p_{\mbox{\tiny{ren}}}(t) \, = \,
-\frac{\Lambda}{8\pi G} \, - \,
\frac{\lambda H^4}{2^6\pi^4} \Bigg[ \,
\frac12 \ln^2 a \, + \,
\frac13 \ln a \, + \,
\frac16 \sum_{n=1}^{\infty} \frac{n^2 - 4}{(n+1)^2} \, a^{-n-1}
\Bigg] \; + \; O(\lambda^2)
\;\; . \label{pphi4}
\end{equation}
This is a completely renormalized result that -- besides the leading
infrared term -- explicitly exhibits all the sub-leading infrared
pieces. The one-loop effect from the kinematic energies of
inflationary scalars is again a constant that must be subsumed into
$\Lambda$. The next order effect is secular because the $\varphi^4$
self-interaction involves the total scalar field strength which grows
as more and more scalars are produced. The induced energy density is
positive, for $\lambda > 0$, because the $\varphi^4$ term adds to the
energy density. In this model we might also guess that the effect is
self-limiting because the classical restoring force tends to push the
scalar back to towards zero energy density. \\

$\bullet$ {\it Infrared Logarithms:} Both gravitation and the scalar model,
as (\ref{rhoGR}-\ref{pGR}) and (\ref{rhophi4}-\ref{pphi4}) show, exhibit
infrared logarithms -- factors of $\ln(a) = Ht$. 
\footnote{Secular terms are ubiquitous in quantum field theory; for
example see \cite{CGO,BV4}.}
As inflation proceeds, these infrared logarithms grow without bound 
until they eventually overcome the small coupling constants -- 
$(\kappa H)^2$ for gravity and $\lambda$ for the scalar model. We cannot 
conclude that there is actually a significant change in the background 
expansion rate because the higher order results remain unknown. The 
legitimate conclusions are rather that: \\
{\it (i)} The expansion rate decreases for gravitation and increases
for the scalar model; and \\
{\it (ii)} Both effects eventually become non-perturbatively strong. \\
In Section 2 we explain why infrared logarithms occur, both from the
mathematics of perturbation theory and on physical grounds. We also
show that the field operator behaves like a stochastic random variable
in the leading logarithm approximation. \\

$\bullet$ {\it Non-perturbative Extension:} Since time evolution makes
the secular growth of the infrared logarithms unbounded, perturbation
theory eventually breaks down. To reliably find out what happens
after the breakdown, we must develop a non-perturbative technique.
\footnote{For non-perturbative methods different from the one we shall
describe here see \cite{BV1}.}
This seemingly impossible task may have a reasonable chance to produce
a satisfactory method because inflationary evolution eventually makes
the part of the quantum field responsible for particle production
stochastic. Since it is precisely this part of the full field that
drives the infrared effects of interest, we can try to isolate its
contribution in the equations of motion and solve for its evolution
non-perturbatively.

Starobinski\u{\i} has long argued that his technique of stochastic
inflation \cite{starobinsky2} should recover the leading infrared logarithms
at each order in perturbation theory. A known example where the
leading infrared late time evolution has been calculated beyond
perturbation theory using stochastic techniques, is the scalar model
(\ref{Lphi4}); this was done by Starobinski\u{\i} and Yokoyama
\cite{starobinsky3} and its explication occupies Section 4. Their result 
agrees with the intuitive expectation that the growth of its field 
strength is eventually halted by the classical restoring force of the 
potential.

The gravitational system (\ref{Lgr}) to which we should like to apply
the same method, differs in many ways from the particular scalar model
(\ref{Lphi4}). Two basic differences are the presence of derivative
interactions and constrained fields in the effective gravity theory.
Both differences must be addressed and the stochastic approximation
rules must be extended so that they can be applied to (\ref{Lgr}).
This is done in Section 3 where we also show that the field operator
behaves like a stochastic random variable in the leading logarithm
approximation. In Sections 5 and 6, we consider and analyze suitable
scalar models with derivative interactions and a reasonable analogue 
of constrained fields. Our conclusions comprise Section 7.

The underlying idea behind the stochastic analysis of such theories is
that the quantum field consists of two parts; a part which contains
the ultraviolet effects that just redefine the parameters of the low
energy interactions, and a part which contains the infrared effects from
inflationary particle production and is responsible for the secular
infrared logarithms. The ultraviolet sector decouples from the infrared
except for: \\
{\it (i)} The constant renormalizations of the low energy parameters
its interacting component furnishes; and \\
{\it (ii)} Its function as a reservoir of stochastic perturbations as
the inflationary redshift pushes more and more modes into the infrared.

\section{The Physics of Infrared Logarithms}

The origin of the infrared logarithms in expressions
(\ref{rhoGR}-\ref{pGR}) and (\ref{rhophi4}-\ref{pphi4}) can be
understood physically as well as from the mathematics of perturbation
theory. Although the physical understanding is vastly more important
in guiding the generalization we must make, we shall begin by explaining
how perturbative computations are done invariantly in a locally de Sitter
background. Besides revealing the sources of the infrared logarithms,
these techniques provide an invariant separation between infrared and
ultraviolet degrees of freedom. Moreover, these techniques shall be
used in the computations of Sections 5 and 6 which relate exact results
from perturbation theory to stochastic realizations. \\

$\bullet$ {\it Invariant Regulation:} We employ dimensional regularization
in position space. The vertices are straightforward to obtain in any
background, and the only difficulty comes in finding the propagators in
arbitrary spacetime dimension $D$. These propagators are expressed in
terms of a de Sitter invariant length function we call $y(x;x')$:
\begin{equation}
y(x;x') \; \equiv \; a(t) \, a(t')
\Biggl[ H^2 \Bigl\Vert \vec{x} - \vec{x}' \Bigr\Vert^2
\, - \, \Bigl( -\frac1{a} + \frac1{a'} \Bigr)^2 \Biggr]
\;\; . \label{y}
\end{equation}
Its physical meaning in terms of the invariant length $\ell(x;x')$
between $x^{\mu}$ and $x^{\prime \mu}$ is:
\begin{equation}
y(x;x') = 4 \sin^2\Biggl[ \frac12 H \ell(x;x') \Biggr]
\;\; . \label{y2}
\end{equation}

$\bullet$ {\it Conformal Scalar Propagator:} The simplest propagator is
that of a massless, conformally coupled scalar \cite{birrell}:
\begin{equation}
{i\Delta}_{\mbox{\tiny CF}}(x;x') \; = \;
\frac{H^{D-2}}{(4\pi)^{\frac{D}2}} \;
\Gamma\Bigl( \frac{D}2 \!-\! 1 \Bigr) \;
\Bigl( \frac4{y} \Bigr)^{\frac{D}2-1}
\;\; . \label{propCF}
\end{equation}
Because $y(x;x')$ vanishes when $x^{\prime \mu} = x^{\mu}$, and because
one always interprets $D$ so that zero is raised to a positive power
in dimensional regularization, the coincidence limit of the conformally
coupled propagator vanishes:
\begin{equation}
{i\Delta}_{\mbox{\tiny CF}}(x;x) \; = \; 0
\;\; . \label{propCFcoinc}
\end{equation}

$\bullet$ {\it Scalar Propagator:} Although conformal invariance
suppresses interesting quantum effects, the conformal scalar propagator
is quite useful for expressing the propagator of the minimally coupled
scalar \cite{rpw1,rpw2}:
\begin{eqnarray}
\lefteqn{ i\Delta_A(x;x') \; = \; i\Delta_{\mbox{\tiny CF}}(x;x') }
\label{DeltaA} \\
& & + \,
\frac{H^{D-2}}{(4\pi)^{\frac{D}2}} \,
\frac{\Gamma(D \!-\! 1)}{\Gamma(\frac{D}2)}
\left\{ \frac{D}{D\!-\! 4} \,
\frac{\Gamma^2(\frac{D}2)}{\Gamma(D \!-\! 1)} \,
\Bigl(\frac4{y}\Bigr)^{\frac{D}2 -2} \!\!
- \pi \cot\Bigl( \frac{\pi}2 D \Bigr)
+ \ln(a a') \right\} \nonumber \\
& & +
\frac{H^{D-2}}{(4\pi)^{\frac{D}2}}
\sum_{n=1}^{\infty} \left\{
\frac1{n} \,
\frac{\Gamma(n \!+\! D \!-\! 1)}{\Gamma(n \!+\! \frac{D}2)} \,
\Bigl(\frac{y}4 \Bigr)^n \!\!
- \frac1{n \!-\! \frac{D}2 \!+\! 2} \,
\frac{\Gamma(n \!+\! \frac{D}2 \!+\! 1)}{\Gamma(n \!+\! 2)} \,
\Bigl(\frac{y}4 \Bigr)^{n - \frac{D}2 + 2} \right\}
\;\; . \nonumber
\end{eqnarray}
This expression might seem daunting but it is actually simple to use in
low order computations because the infinite sum on the final line vanishes
in $D=4$, and each term in the series goes like a positive power of $y(x;x')$.
This means that the infinite sum can only contribute when multiplied by a
divergence, and even then only a small number of terms can contribute. \\

$\bullet$ {\it Correlation Source:} The explicit factor of $\ln(a a')$
present in (\ref{DeltaA}) is one source of infrared logarithms. It gives
the secular dependence of the coincidence limit 
\cite{vilenkin,linde,starobinsky4}:
\begin{equation}
{i\Delta}_A(x;x) \; = \;
\frac{H^{D-2}}{(4\pi)^{\frac{D}2}} \,
\frac{\Gamma(D \!-\! 1)}{\Gamma(\frac{D}2)}
\left\{ -\pi \cot\Bigl( \frac{\pi}2 D \Bigr)
\, + \, 2 \ln a \right\}
\; \; . \label{propAcoinc}
\end{equation}
This factor of $\ln(a a')$ is also interesting in that it breaks de Sitter
invariance. Of course $\varphi(x)$ transforms like a scalar; the breaking
derives from the state in which the expectation value of the two free fields
is taken. Allen and Folacci long ago proved that the massless, minimally
coupled scalar fails to possess normalizable de Sitter invariant states
\cite{allen}. A final point is that the factor of $\ln(a a')$ is actually 
an invariant, it just depends upon the initial value surface at which the
state is released. In fact, it is the Hubble constant times the sum of the
invariant times from $x^{\mu}$ and $x^{\prime \mu}$ to this initial value
surface. The physical reason for the appearance of such a term is the
increasing field amplitude due to inflationary particle production. \\

$\bullet$ {\it Graviton Propagator:} We define the graviton field
$\psi_{\mu\nu}(x)$ as follows:
\begin{equation}
g_{\mu\nu}(x) \, \equiv \,
a^2 \Bigl( \eta_{\mu\nu} + \kappa \psi_{\mu\nu}(x) \Bigr)
\qquad , \qquad
\kappa^2 \equiv 16 \pi G
\;\; . \label{psimn}
\end{equation}
Our gauge fixing Lagrangian takes the form:
\begin{equation}
\mathcal{L}_{\mbox{\tiny GF}} \; = \;
-\frac12 \, a^{D-2} \, \eta^{\mu\nu} F_{\mu} F_{\nu}
\;\; , \label{Lgf}
\end{equation}
where $F_{\mu}$ is an analogue of the de Donder gauge fixing term 
of flat space \cite{nctrpw3}:
\begin{equation}
F_{\mu} \; \equiv \;
\eta^{\rho\sigma} \Bigl[ \,
\psi_{\mu\rho , \sigma} \, - \,
\frac12 \psi_{\rho \sigma , \mu} \, + \,
(D \!-\! 2) \, H a \, \psi_{\mu \rho} \; \delta^0_{\sigma}
\; \Bigr)
\;\; . \label{gf}
\end{equation}
Because space and time components are treated differently it is useful
to have an expression for the purely spatial part of the Minkowski
metric:
\begin{equation}
\overline{\eta}_{\mu\nu} \, \equiv \,
\eta_{\mu\nu} + \delta^0_{\mu} \, \delta^0_{\nu}
\;\; . \label{etabar}
\end{equation}
With these definitions the graviton propagator takes the form of a
sum of three constant index factors times three scalar propagators:
\begin{equation}
i\Bigl[ {}_{\mu\nu} \Delta_{\rho\sigma} \Bigr](x;x') \, = \,
\sum_{I=A,B,C} \,
\Bigl[ {}_{\mu\nu} T^I_{\rho\sigma} \Bigr] \, i\Delta_I(x;x')
\;\; . \label{gprop}
\end{equation}
The explicit expressions for the index factors are \cite{rpw3}:
\begin{eqnarray}
\Bigl[ {}_{\mu\nu} T^A_{\rho\sigma} \Bigr] & = &
2 \, \overline{\eta}_{\mu (\rho} \, \overline{\eta}_{\sigma) \nu}
\, - \, \frac2{D\!-\! 3} \,
\overline{\eta}_{\mu\nu} \, \overline{\eta}_{\rho \sigma}
\;\; , \label{T_A} \\
\Bigl[ {}_{\mu\nu} T^B_{\rho\sigma} \Bigr] & = &
-4 \, \delta^0_{(\mu} \,
\overline{\eta}_{\nu) (\rho} \, \delta^0_{\sigma)}
\;\; , \label{T_B} \\
\Bigl[ {}_{\mu\nu} T^C_{\rho\sigma} \Bigr] & = &
\frac2{(D \!-\! 2) (D \!-\! 3)} \,
\Bigl[ \, (D \!-\! 3) \, \delta^0_{\mu} \, \delta^0_{\nu} \, + \,
\overline{\eta}_{\mu\nu} \, \Bigr]
\Bigl[ \, (D \!-\! 3) \, \delta^0_{\rho} \, \delta^0_{\sigma} \, + \,
\overline{\eta}_{\rho \sigma} \, \Bigr]
 \qquad \label{T_C}
\end{eqnarray}
The $A$-type propagator is identical to (\ref{DeltaA}), while the
$B$-type and $C$-type are given by:
\begin{eqnarray}
i \Delta_B(x;x') & = & i \Delta_{\mbox{\tiny CF}}(x;x') \; - \;
\frac{H^{D-2}}{(4\pi)^{\frac{D}2}} \,
\sum_{n=0}^{\infty} \Bigg\{
\frac{\Gamma(n \!+\! D \!-\! 2)}{\Gamma(n \!+\! \frac{D}2)} \,
\Bigl( \frac{y}4 \Bigr)^n
\nonumber \\
& \mbox{} & \qquad\qquad\qquad
- \, \frac{\Gamma(n \!+\! \frac{D}2)}{\Gamma(n \!+\! 2)} \,
\Bigl( \frac{y}4 \Bigr)^{n - \frac{D}2 +2} \Bigg\}
\;\; , \label{DeltaB} \\
i \Delta_C(x;x') & = & i \Delta_{\mbox{\tiny CF}}(x;x') \; + \;
\frac{H^{D-2}}{(4\pi)^{\frac{D}2}} \,
\sum_{n=0}^{\infty} \Bigg\{ (n \!+\! 1) \,
\frac{\Gamma(n \!+\! D \!-\! 3)}{\Gamma(n \!+\! \frac{D}2)} \,
\Bigl( \frac{y}4 \Bigr)^n
\nonumber \\
& \mbox{} & \qquad\qquad\qquad
- \Bigl( n \!-\! \frac{D}2 \!+\! 3 \Bigr) \,
\frac{\Gamma(n \!+\! \frac{D}2 \!-\! 1)}{\Gamma(n \!+\! 2)} \,
\Bigl( \frac{y}4 \Bigr)^{n - \frac{D}2 + 2} \Bigg\}
\;\; . \qquad \label{DeltaC}
\end{eqnarray}
For completeness we also give the ghost propagator in this gauge:
\begin{equation}
i\Bigl[ {}_{\mu} \Delta_{\nu} \Bigr](x;x') \; = \;
\overline{\eta}_{\mu\nu} \; i\Delta_A(x;x') \; - \;
\delta^0_{\mu} \, \delta^0_{\nu} \, i\Delta_B(x;x')
\;\; . \label{Deltagh}
\end{equation}
Note that the infinite sums in (\ref{DeltaB}) and (\ref{DeltaC})
vanish for $D=4$ so that, in this limit, the $B$-type and $C$ type
propagators both agree with the conformal propagator. It is significant
that only the $A$-type propagator contributes infrared logarithms.
Since only physical graviton modes can experience the inflationary
particle production responsible for infrared logarithms, we expect
only the $A$-type propagator to contain them and this is indeed the
case. A final point is that the breaking of de Sitter invariance
apparent in these infrared logarithms is a real effect, not an artifact
of having employed a de Sitter non-invariant gauge. One way to prove
this is by defining a de Sitter transformation of the graviton field
to include the compensating diffeomorphism needed to restore the gauge
condition \cite{kleppe}. \\

$\bullet$ {\it Volume Source:} Although any infrared logarithms which
appear in one-loop diagrams can only have come from the $A$-type propagator,
there is another mechanism that can produce infrared logarithms in higher
loop results such as (\ref{rhoGR}-\ref{pGR}) and
(\ref{rhophi4}-\ref{pphi4}). This other mechanism is the growth of
the invariant volume of the past light-cone from the observation point
back to the initial value surface:
\begin{eqnarray}
V_{\mbox{\tiny PLC}}(t) & \equiv &
\int_0^t dt' \, a^{\prime {\scriptscriptstyle D-1}}
\int d^{\scriptscriptstyle D-1}x' \,\,
\theta\Bigl[-y(x;x')\Bigr]
\;\; , \label{Vplc} \\
& = &
\frac{2 \pi^{\frac{D-1}2} H^{-D}}{(D \!-\! 1) \, \Gamma(\frac{D-1}2)} \,
\Biggl[ \, \ln a + O(1) \, \Biggr]
\;\; . \label{Vplclimit}
\end{eqnarray}
Factors of this quantity arise naturally whenever undifferentiated
propagators connect an interaction vertex -- at $x^{\prime \mu}$ --
with the expectation value of some observable -- at $x^{\mu}$.

To obtain true expectation values for cases -- such as cosmology --
in which the {\it ``in''} and {\it ``out''} vacua either do not agree
or are not even well defined \cite{jordan}, the Schwinger-Keldysh formalism
\cite{schwinger} must be employed. In this formalism the only net corrections
come from interaction vertices which lie on or within the past light-cone
of some observation point. The more familiar {\it ``in-out''} matrix
elements would harbor virulent infrared divergences from integrating
over the exponentially large inflationary future volume \cite{nctrpw4}.
The causality of the Schwinger-Keldysh formalism regulates these infrared
divergences but simple correspondence implies that the regulated
expressions must grow without bound at late times.

\section{Infrared Dynamics and Their Rules}

A crucially important consequence of our use of an invariant
ultraviolet regularization is that there should be a constant
dynamical impact from all modes whose physical wavelength ranges
from zero -- the far ultraviolet -- to any fixed value. Recall
that quanta are labeled by constant wave vectors $\vec{k}$, and
that the mode with wavenumber $k = \Vert \vec{k} \Vert$ begins
to experience significant inflationary particle production when
the number of particles $N(t, k) > 1$ or, equivalently, when
$k < H a(t)$. This suggests the following physical separation
between ``infrared'' and ``ultraviolet'' modes:
\begin{eqnarray}
{\it Infrared} \qquad & \Longrightarrow & \qquad H < k < H a(t)
\;\; , \label{irmodes} \\
{\it Ultraviolet}
\qquad & \Longrightarrow & \qquad k > H a(t)
\;\; . \label{uvmodes}
\end{eqnarray}
The range of the physical wavelengths $\lambda_{\rm ph} =
2\pi k^{-1} a(t)$ of the ultraviolet modes is from zero to the
constant $H^{-1}$. With any invariant regularization, the dynamical
impact of such modes must be constant because they lie in an
invariantly defined range.

Had we taken the lower limit in expression (\ref{irmodes}) down
to $k=0$, the infrared phase space would also have extended over a
constant physical range. However, taking $k$ down to zero has long
been known to result in infrared divergences \cite{ford2}. We regulate
these by working on the manifold $T^{D-1} \times \Re$, with the range
of the toroidal coordinates equal to a Hubble length \cite{nctrpw4}:
\begin{equation}
- \frac12 H^{-1} < x^i \leq \frac12 H^{-1}
\;\; . \label{T(D-1)}
\end{equation}
Other regulating techniques exist \cite{fulling} but they all cause the
effective infrared phase space to increase as the universe inflates.
Indeed, this growth is the physical source of infrared logarithms.
What happens is that the average field strength increases as it receives
contributions from more and more infrared modes. If any interactions
involve the undifferentiated field, then this growth can enter into
physical quantities. \\

$\bullet$ {\it Self-interacting Scalar:} These assertions can be confirmed
in the context of a self-interacting scalar theory in $D = 3 + 1$ dimensions
and in the presence of the inflationary background (\ref{dS}):
\begin{equation}
{\cal L} \; = \;
- \frac12 \sqrt{-g} \; g^{\mu\nu} \;
\partial_{\mu} \varphi \; \partial_{\nu} \varphi
\; - \; \sqrt{-g} \; V(\varphi)
\;\; , \label{Lmms}
\end{equation}
where, for stability reasons, the potential $V(\varphi)$ is to be bounded
from below. The resulting field equation is:
\begin{equation}
\ddot{\varphi} \, + \, 3 H \, \dot{\varphi} \, - \,
\frac{1}{a^2} \; \nabla^2 \varphi \, + \, V'(\varphi)
\; = \; 0
\;\; . \label{fulleqn}
\end{equation}
The full -- ultraviolet plus infrared -- perturbative initial value
solution can be obtained by iterating a Yang-Feldman equation \cite{YF}
for which the ``{\it in}'' time has been set to $t=0$:
\begin{equation}
\varphi(t,\vec{x}) \; = \;
\varphi_0(t,\vec{x}) \, - \,
\int_0^t dt' \; a^3(t')
\int d^3x' \; G_{\rm ret}\Bigl( t,\vec{x} \ ; t',\vec{x}' \Bigr) \;
V'(\varphi)(t',\vec{x}')
\;\; . \label{perteqn}
\end{equation}
The retarded Green's function equals:
\begin{eqnarray}
\hspace{-1cm}
G_{\rm ret}\Bigl( t,\vec{x} \ ; t',\vec{x}' \Bigr) & \equiv &
\frac{H^2}{4\pi} \; \theta(t - t') \,
\Bigg\{ \, \frac{\delta\Bigl( H \Vert\vec{x} - \vec{x}'\Vert +
\frac1{a} - \frac1{a'} \Bigr)} {a a' \, H \Vert\vec{x} - \vec{x}'\Vert}
\nonumber \\
& \mbox{} & \hspace{3cm}
+ \ \theta\Bigl( H \Vert\vec{x} - \vec{x}'\Vert +
\frac1{a} - \frac1{a'} \Bigr) \Bigg\}
\;\; . \label{Gret}
\end{eqnarray}
The free field $\varphi_0(t,\vec{x})$ is expanded in terms of mode 
functions $u(t,k)$ and operators $\alpha(\vec{k})$ and 
$\alpha^{\dagger}(\vec{k})$ obeying canonical commutation relations:
\begin{equation}
\varphi_0(t,\vec{x}) \, = \,
\int \frac{d^3k}{(2\pi)^3} \;
\Biggl\{ e^{i \vec{k} \cdot \vec{x}} \, u(t,k) \; \alpha(\vec{k}) \, + \,
e^{-i \vec{k} \cdot \vec{x}} \, u^*(t,k) \; \alpha^{\dagger}(\vec{k})
\Biggr\}
\;\; , \label{freefield}
\end{equation}
where:
\begin{eqnarray}
u(t,k) & = &
\frac{H}{\sqrt{2 k^3}} \,
\Biggl[ \, 1 \, - \, \frac{i k}{H a} \, \Biggr] \;
e^{\frac{ik}{Ha}}
\;\; , \label{modefncts} \\
\Bigl[ \, \alpha(\vec{k}) \, , \, \alpha^{\dagger}(\vec{k}') \, \Bigr]
& = &
(2\pi)^3 \, \delta^3\Bigl( \vec{k} - \vec{k}' \Bigr)
\;\; . \label{ccr}
\end{eqnarray}
It should be noted that although $\varphi_0(t,\vec{x})$ is only the
lowest order part of the solution, it and its first derivative agree
exactly with the full field on the initial value surface. \\

$\bullet$ {\it Infrared Field:} To excise the ultraviolet modes
(\ref{uvmodes}), we iterate what is essentially the same equation:
\begin{equation}
\Phi(t,\vec{x}) \; = \;
\Phi_0(t,\vec{x}) \, - \,
\int_0^t dt' \; a^3(t')
\int d^3x' \; G_{\rm ret}\Bigl( t,\vec{x} \ ; t',\vec{x}' \Bigr) \;
V'(\Phi)(t',\vec{x}')
\; , \label{fullIR}
\end{equation}
but with the zeroth order solution restricted to only infrared modes:
\begin{equation}
\Phi_0(t,\vec{x}) =
\int \frac{d^3k}{(2\pi)^3} \; \theta\Bigl( Ha(t) - k \Bigr)
\Biggl\{ e^{i \vec{k} \cdot \vec{x}} \, u(t, k) \; \alpha_{\vec{k}}
\, +
e^{-i \vec{k} \cdot \vec{x}} \, u^*(t, k) \; \alpha^{\dagger}_{\vec{k}}
\Biggr\}
\; . \label{freefieldIR}
\end{equation}

This model is completely free of ultraviolet divergences, so we are
justified in taking $D = 3 + 1$. To see that the model also reproduces
the leading infrared logarithms at tree order it suffices to take the
vacuum expectation value of $\Phi_0^2$:
\footnote{Vacuum expectation values are taken throughout in the presence
of the free Bunch--Davies vacuum at $t=0$.}
\begin{eqnarray}
\Big\langle \;
\Phi^2_0(t,\vec{x})
\; \Big\rangle_{\mbox{\tiny VEV}}
& = &
\int \frac{d^3k}{(2\pi)^3} \;
\theta\Bigl( H a - k \Bigr) \, \frac{H^2}{2 k^3} \,
\Bigg\{ 1 + \frac{k^2}{H^2 a^2} \Biggr\}
\;\; , \\
& = &
\frac{H^2}{4 \pi^2} \int_H^{Ha} \frac{dk}{k} \;
\Bigg\{ 1 + \frac{k^2}{H^2 a^2} \Biggr\}
\;\; , \\
& = &
\Biggl( \frac{H}{2\pi} \Biggr)^2 \,
\Biggl\{ \ln a + \frac12 - \frac1{2 a^2} \Biggr\}
\;\; . \label{Phi0sq}
\end{eqnarray}
Comparison with (\ref{propAcoinc}) for $D = 3+1$ reveals exact
agreement between the $\ln(a)$ terms. \\

$\bullet$ {\it Infrared Field Equation:} Our purely infrared field operator
$\Phi(t,\vec{x})$ does not quite obey the original field equation 
(\ref{fulleqn}) because the kinetic operator fails to annihilate 
$\Phi_0(t,\vec{x})$. The reason is that, upon taking time derivatives of 
$\Phi_0(t,\vec{x})$, there are extra contributions due to the presence of 
the time dependent upper limit $H a(t)$ of the mode sum in (\ref{freefieldIR}).
Thus, the action of the kinetic operator on $\Phi_0(t,\vec{x})$ produces 
momentum space surface terms:
\begin{equation}
\ddot{\Phi}_0(t,\vec{x}) + 3 H \dot{\Phi}_0(t,\vec{x}) -
\frac{\nabla^2}{a^2(t)} \Phi_0(t,\vec{x}) \, = \,
\dot{\mathcal{F}}(t,\vec{x}) + \mathcal{G}(t,\vec{x}) +
3 H \mathcal{F}(t,\vec{x})
\;\; , \label{extra}
\end{equation}
involving the sources:
\begin{eqnarray}
\mathcal{F}(t,\vec{x}) \!\!& \equiv &\!\!
H \!\!\int\! \frac{d^3k}{(2\pi)^3} \, k \,
\delta\Bigl( k \!-\! H a(t) \Bigr)
\Biggl\{ e^{i \vec{k} \cdot \vec{x}} \, u(t,k) \, \alpha_{\vec{k}}
\, +
e^{-i \vec{k} \cdot \vec{x}} \, u^*(t,k) \, \alpha^{\dagger}_{\vec{k}}
\Biggr\}
\qquad \label{F} \\
\mathcal{G}(t,\vec{x}) \!\!& \equiv &\!\!
H \!\!\int\! \frac{d^3k}{(2\pi)^3} \, k \,
\delta\Bigl( k \!-\! H a(t) \Bigr)
\Biggl\{ e^{i \vec{k} \cdot \vec{x}} \, \dot{u}(t,k) \, \alpha_{\vec{k}}
\, +
e^{-i \vec{k} \cdot \vec{x}} \, \dot{u}^*(t,k) \, \alpha^{\dagger}_{\vec{k}}
\Biggr\}
\label{G}
\end{eqnarray}
Note that the mode functions and their derivatives are simply constants
at $k = H a(t)$:
\begin{eqnarray}
u(t,k) \Biggl\vert_{k = H a(t)} =
\frac{H}{\sqrt{2 k^3}} \, \Bigl( 1 - i \Bigr) e^i
\quad & \Longrightarrow & \quad
\vert u(t,k) \vert^2 \, \Biggl\vert_{k = H a(t)} =
\frac{H^2}{k^3}
\;\; , \qquad \label{hormodes} \\
\dot{u}(t,k) \Biggl\vert_{k = H a(t)} =
-\frac{H^2}{\sqrt{2 k^3}} \, e^i
\hspace{0.95cm} \quad & \Longrightarrow & \quad
\vert \dot{u}(t,k) \vert^2 \, \Biggl\vert_{k = H a(t)} =
\frac{H^4}{2 k^3}
\;\; . \label{hormodesder}
\end{eqnarray}

In view of (\ref{extra}) the equation obeyed by the infrared field
$\Phi(t,\vec{x})$ is not (\ref{fulleqn}) but rather:
\begin{equation}
\Bigl( \ddot{\Phi} - \dot{\mathcal{F}} - \mathcal{G} \Bigr)
\, + \, 3 H \Bigl( \dot{\Phi} - \mathcal{F} \Bigr)
\, - \, \frac{\nabla^2}{a^2} \Phi
\, + \, V'(\Phi)
\; = \; 0
\;\; . \label{Phieqn}
\end{equation}

$\bullet$ {\it Infrared Conservation:} The infrared field equation
(\ref{Phieqn}) does not leave the original stress-energy tensor conserved
because stress-energy is being continually dumped into the truncated system
by ultraviolet modes which redshift past the horizon. We can account for
this by modifying what we call $T_{\mu\nu}$. To motivate the modification
it is useful to write down the form we expect for the divergence of the
stress-energy. For the zero component we should get:
\begin{eqnarray}
T_{0 \mu}^{~~;\mu} & = &
- \dot{T}_{00} - H \Bigl[ 3 T_{00} + g^{ij} T_{ij} \Bigr]
+ g^{ij} T_{0i,j}
\;\; , \\
& = &
- \dot{\Phi} \Biggl[
\Bigl( \ddot{\Phi} - \dot{\mathcal{F}} - \mathcal{G} \Bigr)
+ 3 H \Bigl( \dot{\Phi} - \mathcal{F} \Bigr)
- \frac{\nabla^2}{a^2} \Phi + V'(\Phi) \Biggr]
\;\; ,\label{timeT}
\end{eqnarray}
while for the spatial components:
\begin{eqnarray}
T_{i \mu}^{~~;\mu} & = &
- \dot{T}_{0i} - 3 H T_{0i} + \frac1{a^2} T_{ij,j}
\;\; , \\
& = &
- \partial_i \Phi \Biggl[
\Bigl( \ddot{\Phi} - \dot{\mathcal{F}} - \mathcal{G} \Bigr)
+ 3 H \Bigl( \dot{\Phi} - \mathcal{F} \Bigr)
- \frac{\nabla^2}{a^2} \Phi + V'(\Phi) \Biggr]
\;\; . \label{spaceT}
\end{eqnarray}

We can enforce (\ref{timeT}) and (\ref{spaceT}) with a stress-energy
of the form:
\begin{eqnarray}
T_{00} & = &
\frac12 \Bigl( \dot{\Phi} \!-\! \mathcal{F} \Bigr)^2 +
\frac1{2a^2} \vec{\nabla} \Phi \!\cdot\! \vec{\nabla} \Phi +
V(\Phi)
\;\; , \label{T_00} \\
T_{0i} & = &
\Bigl( \dot{\Phi} \!-\! \mathcal{F} \Bigr) \, \partial_i\Phi
\;\; , \label{T_0i} \\
T_{ij} & = &
\partial_i \Phi \; \partial_j \Phi -
g_{ij} \Biggl[ -\frac12 \Bigl( \dot{\Phi} - \mathcal{F} \Bigr)^2 +
\frac1{2 a^2} \vec{\nabla} \Phi \!\cdot\! \vec{\nabla} \Phi +
V(\Phi) \Biggr] \nonumber \\
& &
+ \, \partial_i S_j + \partial_j S_i -
\frac12 \Bigl( \delta_{ij} \!-\! 3\frac{\partial_i \partial_j}{\nabla^2}
\Bigr) S^L +
\frac12 \Bigl( \delta_{ij} \!-\! \frac{\partial_i \partial_j}{\nabla^2}
\Bigr) S
\;\; . \label{Tconc}
\end{eqnarray}
Once this form is assumed, the non-local source terms of the purely
spatial components are determined by conservation:
\begin{eqnarray}
S_i \!\!& = &\!\!
\frac{a^2}{\nabla^2} \Biggl[
\mathcal{G} \, \partial_i \Phi +
( \dot{\Phi} \!-\! \mathcal{F} ) \, \partial_i \mathcal{F} \Biggr]
- \frac{a^4 \, \partial_i \partial_k}{\nabla^4} \Biggl[
\mathcal{G} \, \partial_k \Phi +
( \dot{\Phi} \!-\! \mathcal{F} ) \, \partial_k \mathcal{F} \Biggr]
\;\; , \\
S^L \!\!& = &\!\!
\frac{a^2 \, \partial_k}{\nabla^2} \Biggl[
\mathcal{G} \, \partial_k \Phi +
( \dot{\Phi} \!-\! \mathcal{F} ) \, \partial_k \mathcal{F} \Biggr]
\;\; , \\
S \!\!& = &\!\!
\frac{a^2}{H} \mathcal{F} \Biggl[
\ddot{\Phi} \!-\! \dot{\mathcal{F}} \!-\! \mathcal{G}
\!+\! 3H ( \dot{\Phi} \!-\! \mathcal{F} )
\!-\! \frac{\nabla^2}{a^2} \Phi \Biggr]
- \frac{a^2}{H} \Bigl( \dot{\Phi} \!-\! \mathcal{F} \Bigr) \mathcal{G}
- H \vec{\nabla} \mathcal{F} \!\cdot\! \vec{\nabla} \Phi
\qquad \\
\!\!& = &\!\!
- \frac{a^2}{H} \Bigl( \dot{\Phi} \!-\! \mathcal{F} \Bigr) \mathcal{G}
- \frac1{H} \vec{\nabla} \mathcal{F} \!\cdot\! \vec{\nabla} \Phi
- \frac{a^2}{H} \mathcal{F} \, V'(\Phi)
\;\; . \label{concS}
\end{eqnarray}
We have, therefore, achieved a completely consistent model of just the
infrared modes that reproduces the leading infrared logarithms. Note
that this establishes the decoupling of the ultraviolet sector. \\

$\bullet$ {\it Leading Infrared Field:} Although the zeroth order field
$\Phi_0(t,\vec{x})$ contains only infrared modes, it is still quantum
mechanical in that the field and its first time derivative do not
commute:
\begin{eqnarray}
\Bigl[ \Phi_0(t,\vec{x}) \, , \dot{\Phi}_0(t,\vec{x}') \Bigr]
& \!\!=\!\! &
\int \frac{d^3k}{(2 \pi)^3} \;
\theta\Bigl( H a(t) \!-\! k \Bigr)
\Bigl[ \, u \, \dot{u}^* \, - \, u^* \, \dot{u} \, \Bigr]
e^{i \vec{k} \cdot \Delta \vec{x}}
\;\; , \\
& =\!\! &
\frac{i}{4 \pi^2 a^3}
\int_H^{Ha}  dk \; k^2 \,
\frac{\sin(k \Delta x)}{k \Delta x}
\;\; , \\
& =\!\! &
\frac{i}{4 \pi^2 (a \Delta x)^3} \left\{
\matrix{\sin(a H \Delta x) - a H \Delta x \cos(a H \Delta x) \cr
- \sin(H \Delta x) + H \Delta x \cos(H \Delta x)}
\right\}
\qquad \label{Phiccr}
\end{eqnarray}
where $\Delta \vec{x} \equiv \vec{x} - \vec{x}'$ and
$\Delta x \equiv \Vert \Delta \vec{x} \Vert$. However, the leading
infrared logarithm in (\ref{Phi0sq}) derives entirely from the constant
first term of the long wavelength expansion of the mode function:
\begin{eqnarray}
u(t,k) & = &
\frac{H}{\sqrt{2 k^3}} \,
\Biggl[ \, 1 \, - \, \frac{i k}{H a} \, \Biggr] \;
e^{\frac{ik}{Ha}}
\;\; , \\
& = &
\frac{H}{\sqrt{2k^3}} \, \Biggl\{
1 + \frac12 \Bigl(\frac{k}{Ha}\Bigl)^2 + \,
\frac{i}3 \Bigl(\frac{k}{Ha}\Bigl)^3 + \, O(k^4)
\Biggr\}
\;\; . \label{modefnctsexp}
\end{eqnarray}
By constructing a free field, which we shall call $\phi_0(t, {\vec x})$,
with only this first term as its mode function:
\begin{equation}
{\it Leading \; Infrared}
\qquad \Longrightarrow \qquad
u_{\mbox{\tiny IR}}(t, k) \, \sim \,
\frac{H}{\sqrt{2k^3}}
\;\; , \label{irmodefncts} \\
\end{equation}
we would get precisely the same leading infrared logarithm:
\begin{equation}
\phi_0(t,\vec{x}) \equiv
\int \frac{d^3k}{(2\pi)^3} \;
\theta\Bigl( Ha(t) - k \Bigr) \,
\frac{H}{\sqrt{2 k^3}} \, \Biggl\{ \,
e^{i \vec{k} \cdot \vec{x}} \, \alpha_{\vec{k}} \, + \,
e^{-i \vec{k} \cdot \vec{x}} \, \alpha^{\dagger}_{\vec{k}}
\, \Biggr\}
\;\; . \label{phiLIR}
\end{equation}
Because -- unlike a usual quantum field -- the creation and
annihilation parts of $\phi_0(t, {\vec x})$ are both multiplied
by a phase factor with identical time dependence, the leading
infrared field commutes with its time derivative:
\begin{equation}
\left[ \,
\phi_0(t, {\vec x}) \; , \,
\phi_0(t', {\vec x}^{\, \prime})
\, \right]
\; = \; 0
\;\; . \label{phiLIRccr}
\end{equation}
Consequently, $\phi_0$ behaves like a classical variable except for
the operators $\alpha$ and $\alpha^{\dagger}$ which can take
random values. Such a random but commuting variable might be termed 
``stochastic''. \\

$\bullet$ {\it Leading Infrared Field Equation:} Replacing $\Phi_0$
with $\phi_0$ in the basic equation (\ref{fullIR}) does capture the
leading infrared logarithms, but the resulting field theory is still
more complicated than necessary. The leading infrared logarithms are
preserved if we retain only the first term in the long wavelength
expansion of the retarded Green's function. From the expression:
\begin{eqnarray}
G_{\rm ret} \Bigl( t,\vec{x} \ ; t',\vec{x}' \Bigr)
& = &
i \theta(t-t') \times
\label{fullGret} \\
& &
\int \frac{d^3k}{(2\pi)^3} \,
\Bigl[ \, u(t,k) \; u^*(t',k) - u^*(t,k) \; u(t',k) \, \Bigr] \,
e^{i \vec{k} \cdot \Delta \vec{x}}
\;\; , \quad \nonumber
\end{eqnarray}
we can determine the long wavelength expansion of the bracketed term
using (\ref{modefnctsexp}):
\begin{equation}
\Bigl[ \, u(t,k) \; u^*(t',k) - u^*(t,k) \; u(t',k) \, \Bigr] \, = \,
\frac{i}{3H} \Biggl( \frac1{a^{\prime 3}} - \frac1{a^3} \Biggr)
+ O(k^2)
\;\; . \label{uu*exp}
\end{equation}
Retaining only the leading term from (\ref{uu*exp}) in
(\ref{fullGret}) we conclude:
\begin{equation}
G_{\rm ret} \Bigl( t,\vec{x} \ ; t',\vec{x}' \Bigr)
\; \longrightarrow \;
\frac1{3H} \, \theta(t-t') \; \delta^3(\vec{x} - \vec{x}') \,
\Biggl( \frac1{a^{\prime 3}} - \frac1{a^3} \Biggr)
\;\; . \label{irGret}
\end{equation}
Recall that the Green's function is multiplied by a factor of 
$a^{\prime 3}$ from the measure of integration in the Yang-Feldman
equation. Whereas the term $(a'/a')^3 = 1$ contributes coherently
over the full range of integration, the term $-(a'/a)^3$ is only 
significant for $t' \sim t$. Hence, the second term in (\ref{irGret})
is irrelevant in the leading logarithm approximation:
\begin{equation}
{\it Leading \; Infrared}
\quad \Longrightarrow \quad
a^{\prime 3} \; G_{\rm ret}^{\hspace{0.05cm} \mbox{\tiny IR}}
\Bigl( t,\vec{x} \ ; t',\vec{x}' \Bigr)
\, \sim \,
\frac1{3H} \, \theta(t-t') \; \delta^3(\vec{x} - \vec{x}')
\;\; . \qquad \label{GretLIR}
\end{equation}
We are at length lead to the following simple iterative equation for 
recovering the leading infrared logarithms:
\begin{equation}
\phi(t, \vec{x}) \; = \;
\phi_0(t, \vec{x}) \, - \,
\frac{1}{3H} \int_0^t dt' \;
V'\Bigl( \phi(t',\vec{x}) \Bigr)
\;\; . \label{LIReqn}
\end{equation}

To infer the local differential equation which the full stochastic field
$\phi$ obeys, we note that the time derivative of the stochastic free
field $\phi_0$ is a momentum space surface term:
\begin{eqnarray}
\dot{\phi}_0(t,\vec{x})
& = &
\int \frac{d^3k}{(2\pi)^3} \;
\delta\Bigl( k \!-\! H a(t) \Bigr) \;
\frac{H^2}{\sqrt{2k}} \;
\Biggl\{ \, e^{i \vec{k} \cdot \vec{x}} \, \alpha_{\vec{k}} \, + \,
e^{-i \vec{k} \cdot \vec{x}} \, \alpha^{\dagger}_{\vec{k}} \, \Biggr\}
\;\; , \label{phiLIRdot} \\
& \equiv &
f_{\phi}(t,\vec{x})
\;\; . \label{stochsource}
\end{eqnarray}
Taking the time derivative of (\ref{LIReqn}) gives the Langevin equation
for the stochastic field $\phi(t,\vec{x})$ first obtained by Starobinski\u{\i}
\cite{starobinsky2,starobinsky3}:
\begin{equation}
\dot{\phi}(t,\vec{x}) \; = \; f_{\phi}(t,\vec{x}) \,
- \, \frac1{3H} V'\Bigl( \phi(t,\vec{x})\Bigr) \;\; . \label{stocheqn}
\end{equation}
Physically, $f_{\phi}$ is a stochastic source caused by the ultraviolet 
modes that are instantaneously redshifting across the horizon at time $t$.
Because each increment is uncorrelated, this source represents Gaussian 
white noise:
\begin{equation}
\Big\langle \;
f_{\varphi}(t,\vec{x}) \; f_{\varphi}(t',\vec{x}) 
\; \Big\rangle_{\mbox{\tiny VEV}}
\; = \;
\frac{H^3}{4 \pi^2} \; \delta(t - t') 
\;\; . \label{white}
\end{equation}
The sort of system comprised by (\ref{stochsource}-\ref{stocheqn}) has 
been much studied \cite{volovich}. Expectation values of functionals of 
the stochastic field can be computed in terms of a probability density
$\varrho(t,\phi)$ as follows:
\begin{equation}
\Big\langle \;
F\left[ \, \phi (t, {\vec x}) \right]
\; \Big\rangle_{\mbox{\tiny VEV}}
\; = \;
\int_{-\infty}^{+\infty}
d\omega \; \varrho(t, \omega) \; F(\omega)
\;\; . \label{defdistr}
\end{equation}
The probability density satisfies a Fokker-Planck equation whose first 
term is given by the interaction on the right hand side of equation
(\ref{stocheqn}) and whose second term is fixed by the normalization 
of the white noise (\ref{white}):
\begin{equation}
\dot{\varrho}(t, \phi) \; = \;
\frac{1}{3H} \,
\frac{\partial}{\partial \phi}
\Bigl[ \, V'(\phi) \, \varrho(t, \phi) \, \Bigr] \; + \:
\frac{H^3}{8 \pi^2} \,
\frac{\partial^2}{\partial \phi^2} \Bigl[ \, \varrho(t, \phi) \, \Bigr]
\;\; . \label{fopleqn}
\end{equation}

$\bullet$ {\it Leading Infrared Rules:} Relation (\ref{stocheqn})
derives from the original field equation (\ref{fulleqn}) by applying
the following rules:
\begin{eqnarray}
&\!\!\!\!\! {\rm (I)} &
At \; each \; order \; in \; the \; field \,-\,  \varphi^1, \;
\varphi^2, \; and \; so \; forth \,-\, retain
\nonumber \\
& \mbox{} &
only \; the \; term \; with \; the \; smallest \;
number \; of \; derivatives.
\hspace{2.3cm} \label{ruleI} \\
&\!\!\!\!\! {\rm (II)} &
For \; the \; linear \; terms \; in \; the \; field, \; each \;
time \; derivative \; has
\nonumber \\
& \mbox{} &
a \; stochastic \; source \; subtracted. \label{ruleII}
\end{eqnarray}

For the kinetic part of (\ref{fulleqn}), by applying rule I
(\ref{ruleI}) we select the Hubble friction term and, thereafter,
by applying rule II (\ref{ruleII}) a stochastic source must be
subtracted off because we are at linear order:
\footnote{Since the scale factor varies much faster than the field
during inflation, it is preferable to have derivatives act on the
scale factor instead of the field. Thus, the single time derivative
of the Hubble friction term dominates over the second time and space
derivatives terms.}
\begin{equation}
\ddot{\varphi} \, + \, 3 H \dot{\varphi} \, - \,
\frac{\nabla^2}{a^2} \, \varphi
\quad \longrightarrow \quad
3 H \dot{\varphi}
\quad \longrightarrow \quad
3 H \Bigl( \, \dot{\phi} - f_{\phi} \, \Bigr)
\;\; . \label{KErule}
\end{equation}
There are no derivatives or linear terms in the interaction part of
(\ref{fulleqn}) so we merely replace the full field by its stochastic
counterpart:
\begin{equation}
V'(\varphi)
\quad \longrightarrow \quad
V'(\phi)
\;\; . \label{PErule}
\end{equation}
Hence the full equation (\ref{fulleqn}) reduces as follows,
\begin{equation}
\ddot{\varphi} \, + \, 3 H \dot{\varphi} \, - \,
\frac{\nabla^2}{a^2} \, \varphi + V'(\varphi) \quad \longrightarrow \quad
3 H \Bigl( \, \dot{\phi} - f_{\phi} \, \Bigr) + V'(\phi) \;\; ,
\end{equation}
which is indeed equivalent to (\ref{stocheqn}).

It is important to note that we would get the same leading infrared
logarithms by solving (\ref{Phieqn}) for $\Phi$. The enormous advantage
of solving the stochastic equation (\ref{stocheqn}) instead, is that
the field $\phi$ can be regarded as {\it classical}. Whereas there is
little hope of being able to exactly solve the Heisenberg field equations
for an interacting, four dimensional quantum field, classical field
equations {\it can} sometimes be solved.

\section{Scalar Field with Quartic Self-Interaction}

For the quartic self-interaction (\ref{Lphi4}) the full field equation is:
\begin{equation}
\ddot{\varphi} \, + \, 3H \, \dot{\varphi}
\, - \, \frac{\nabla^2}{a^2} \varphi
\, + \, \frac{\lambda}{6} \, \varphi^3 \; = \; 0
\;\; . \label{phi4eom}
\end{equation}
Applying the two stochastic reduction rules gives:
\begin{equation}
3H \left( \, \dot{\phi} - f_{\phi} \, \right) \, + \,
\frac{\lambda}{6} \, \phi^3
\; = \; 0
\;\; . \label{phi4stocheqn}
\end{equation}

$\bullet$ {\it Non-perturbative Solution:} To compute expectation
values we need the probability density $\varrho(t,\phi)$ which
obeys the Fokker-Planck equation:
\begin{equation}
\frac{\partial}{\partial t} \Bigl[ \varrho(t, \phi) \Bigr] \; = \;
\frac{\lambda}{18H} \,
\frac{\partial}{\partial \phi}
\Bigl[ \phi^3 \, \varrho(t, \phi) \Bigr] \; + \:
\frac{H^3}{8\pi^2} \,
\frac{\partial^2}{\partial \phi^2} \, \Bigl[ \varrho(t, \phi) \Bigr]
\;\; . \label{phi4fopleqn}
\end{equation}
To recover the non-perturbative late time solution of Starobinski\u{\i} 
and Yoko\-yama \cite{starobinsky3} we make the ansatz:
\begin{equation}
\lim_{t \rightarrow \infty} \, \varrho(t, \phi)
\; = \; \varrho_{\infty}(\phi)
\;\; , \label{rhobc}
\end{equation}
because the scalar force should eventually balance the tendency of
inflationary particle production to force the scalar up its potential.
This ansatz results in a first order equation:
\begin{equation}
\frac{d \varrho_{\infty}(\phi)}
{\varrho_{\infty}(\phi)} \; = \;
- \frac{4 \pi^2 \lambda}{9 H^4} \;
\phi^3 d\phi
\;\; . \label{rhoaseqn}
\end{equation}
The solution is straightforward:
\begin{equation}
\varrho_{\infty}(\phi) \; = \;
\frac2{\Gamma(\frac14)} \;
\left( \frac{\pi^2 \lambda}{9H^4} \right)^{\frac14} \;
\exp\left[-\frac{\pi^2}9 \, \lambda \, \Bigl(\frac{\phi}{H}\Bigr)^4 \right]
\;\; , \label{rhoas}
\end{equation}
and its non-perturbative nature is clear. It follows that the stochastic 
expectation value of the $2n$-th power of the field has the following 
late time limit:
\begin{eqnarray}
\lim_{t \rightarrow \infty} \, \Big\langle \;
\phi^{2n}(t,\vec{x})
\; \Big\rangle_{\mbox{\tiny VEV}}
& = &
\int_{-\infty}^{+\infty}
d\omega \; \varrho_{\infty}(\omega) \; \omega^{2n}
\;\; , \\
& = & 
\frac{\Gamma(\frac{n}2 + \frac14)}{\Gamma(\frac14)} \;
\left( \frac{9 H^4}{\pi^2 \lambda} \right)^{\frac{n}2} 
\;\; . \label{phi2n}
\end{eqnarray}

$\bullet$ {\it Perturbative Agreement:} To make contact between stochastic 
expectation values and perturbation theory, we first differentiate the 
vacuum expectation value of $\phi^{2n}$ and then we use the Fokker-Planck 
equation \cite{starobinsky5,FMSVV}:
\begin{eqnarray}
\lefteqn{ \frac{\partial}{\partial t} \left\langle \;
\phi^{2n}(t,\vec{x})
\; \right\rangle_{\mbox{\tiny VEV}} 
=
\int_{-\infty}^{+\infty}
d\omega \; {\dot \varrho}(t, \omega) \; \omega^{2n}
\;\; , } \label{phi4pert} \\
& & =
\frac{n(2n-1) H^3}{4\pi^2} \;
\left\langle \;
\phi^{2n-2}(t,\vec{x})
\; \right\rangle_{\mbox{\tiny VEV}}
\; - \;
\frac{n \lambda}{9H} \;
\left\langle \;
\phi^{2n+2}(t,\vec{x})
\; \right\rangle_{\mbox{\tiny VEV}}
\;\; . \qquad \label{phi4pertsol}
\end{eqnarray}
This relation can be more revealingly expressed in terms of a 
new time variable $\alpha$ and a rescaled coupling constant 
$\overline{\lambda}$:
\begin{equation}
\alpha \, \equiv \,
\frac{1}{4\pi^2} \, \ln a
\qquad , \qquad
\overline{\lambda} \, \equiv \, 
\frac{4\pi^2}{9} \, \lambda
\;\; . \label{rescaled}
\end{equation}
A differential recursion relation emerges:
\begin{equation}
\frac{\partial}{\partial \alpha}  
\left\langle \Bigl( \frac{\phi}{H} \Bigr)^{2n}
\right\rangle_{\mbox{\tiny VEV}} 
= \;
n (2n-1) \left\langle \Bigl( \frac{\phi}{H} \Bigr)^{2n-2} 
\right\rangle_{\mbox{\tiny VEV}} 
- \;
n \overline{\lambda} \left\langle \Bigl( \frac{\phi}{H} \Bigr)^{2n+2} 
\right\rangle_{\mbox{\tiny VEV}} 
\; , \label{recursion}
\end{equation}
whose solution has the form:
\begin{equation}
\left\langle \Bigl( \frac{\phi}{H} \Bigr)^{2n} 
\right\rangle_{\mbox{\tiny VEV}} \; = \; 
(2n-1)!! \;\; \alpha^n \; 
F(\overline{\lambda} \alpha^2) 
\;\; , \label{solution}
\end{equation}
where the function $F_n(z)$ obeys:
\begin{equation}
2 z F_n'(z) + n F_n(z) \, = \,
n F_{n-1}(z) - n (2n+1) z F_{n+1}(z) 
\;\; . \label{Feqn}
\end{equation}
It is straightforward to obtain the first few terms of the series
expansion:
\begin{equation}
F_n(z) \, = \,
1 \, - \,
\frac{n}2 (n+1) z \, + \,
\frac{n}{280} \Bigl(35 n^3 + 170 n^2 + 225 n + 74\Bigr) z^2 
\, + \, O(z^3) 
\;\; . \label{Fexpansion}
\end{equation}
Hence, the stochastic expectation value of $\phi^{2n}$ has the following
time dependence in perturbation theory:
\begin{eqnarray}
\lefteqn{\left\langle \; \phi^{2n}(t,\vec{x})
\; \right\rangle_{\mbox{\tiny VEV}} \; = \;
(2n-1)!! \; \left( \frac{H^2}{4 \pi^2} \ln a \right)^n 
\left\{
1 \, - \, \frac{n}2 (n+1) \; \frac{\lambda}{36 \pi^2} \ln^2 a 
\right.}
\nonumber \\
& & \hspace{1.3cm} \left. 
+ \, \frac{n}{280} \Bigl(35 n^3 + 170 n^2 + 225 n + 74 \Bigr) 
\Bigl[ \, \frac{\lambda}{36 \pi^2} \ln^2 a \, \Bigr]^2 
\, + \, \dots \right\} 
\qquad \label{phi2npert}
\end{eqnarray}
Part of this result is in precise agreement with detailed and highly 
non-trivial calculations of the same quantities using quantum field 
theory \cite{nctrpw5}.

\section{Derivative Interactions}

A basic difference between the scalar model of Section 4 and gravitation,
is the presence of derivative interactions. Since our main purpose is to 
derive the proper stochastic equation for quantum gravity, we first study 
a simple scalar model with derivative interactions defined by the 
following Lagrangian:
\begin{eqnarray}
{\cal L}_{\mbox{\tiny D}} & = &
- \frac12 \, \sqrt{-g} \, g^{\mu\nu} \,
\partial_{\mu} A \; \partial_{\nu} A \;
- \frac12 \, \sqrt{-g} \, g^{\mu\nu} \,
\partial_{\mu} B \; \partial_{\nu} B
\nonumber \\
& \mbox{} &
- \frac{\lambda}{4} \, \sqrt{-g} \, g^{\mu\nu} \,
A^2 \; \partial_{\mu} B \; \partial_{\nu} B
\;\; . \label{Ld}
\end{eqnarray}
Whereas $A$ and $\dot{A}$ have the same equal-time commutation relation
as a free field, $B$ and $\dot{B}$ do not:
\begin{eqnarray}
\Bigl[ \, A(t,\vec{x}) \; , \, \dot{A}(t,\vec{x}') \, \Bigr] 
& = & 
\frac{i \delta^3(\vec{x} - \vec{x}')}{a^3(t)} 
\;\; , \label{can} \\
\Bigl[ \, B(t,\vec{x}) \; , \, \dot{B}(t,\vec{x}') \, \Bigr] 
& = & 
\frac{i \delta^3(\vec{x} - \vec{x}')}
{a^3(t) \, [ \, 1 \, + \, \frac{\lambda}2 A^2(t,\vec{x}) \, ]} 
\;\; . \label{noncan}
\end{eqnarray}
This means we will need to include a homogeneous term in writing the
Yang-Feldman equation for $B$.

The full equations of motion are:
\begin{eqnarray}
\frac1{\sqrt{-g}} \, \frac{\delta S}{\delta A}
& = &
A^{;\mu}_{~~\mu} \, - \,
\frac{\lambda}2 \, g^{\mu\nu} \; A \; B_{,\mu} \; B_{,\nu}
\; = \; 0
\;\; , \label{LdAeom} \\
\frac1{\sqrt{-g}} \, \frac{\delta S}{\delta B}
& = &
B^{;\mu}_{~~\mu} \, + \,
\lambda \, g^{\mu\nu} \, A \, A_{,\mu} \; B_{,\nu} \, + \,
\frac{\lambda}2 \, A^2 \, B^{;\mu}_{~~\mu}
\; = \; 0
\;\; . \label{LdBeom}
\end{eqnarray}
Each equation of motion can be re-written to give the d'Alembertian 
of the field in terms of lower derivatives:
\begin{eqnarray}
A^{;\mu}_{~~\mu} & = & 
\frac{\lambda}2 A \, B_{,\mu} \, B^{,\mu} 
\;\; , \label{AEOM} \\
B^{;\mu}_{~~\mu} & = & 
-\frac{\lambda A \, A_{,\mu} \, B^{,\mu}}{1 + \frac{\lambda}2 A^2} 
\;\; . \label{BEOM}
\end{eqnarray}

The full perturbative initial value solution comes from iterating
the Yang-Feldman equations:\footnote{Field arguments are sometimes 
compressed to 4-vector notation:
$(t, {\vec x}) \equiv x$.}
\begin{eqnarray}
A(t,\vec{x}) & = &
A_0(t,\vec{x}) \, - \,
\frac{\lambda}{2} \int_0^t dt' \; a^3(t')
\int d^3x' \; G_{\rm ret}\Bigl( t,\vec{x} \ ; t',\vec{x}' \Bigr)
\times \label{perteqnAd} \\
& \mbox{} & \hspace{5.5cm}
A(x') \, B_{,\mu}(x') \, B^{,\mu}(x')
\;\; , \nonumber \\
B(t,\vec{x}) & = &
B_0(t,\vec{x}) \, - \,
\frac{\lambda}2 \int d^3x' \; 
G_{\rm ret}\Bigl( t,\vec{x} \ ; 0,\vec{x}' \Bigr)
\; A^2(0,\vec{x}') \, \dot{B}(0,\vec{x}') 
\label{perteqnBd} \\
& \mbox{} & 
+ \, \lambda \int_0^t dt' \; a^3(t')
\int d^3x' \; G_{\rm ret}\Bigl( t,\vec{x} \ ; t',\vec{x}' \Bigr)
\Biggl\{ \frac{ A(x') \, A_{,\mu}(x') \, B^{,\mu}(x')}{1 +
\frac{\lambda}2 A^2(x')} \Biggr\}
\; . \nonumber
\end{eqnarray}
Because both $A$ and $B$ are massless, minimally coupled scalar fields, 
the retarded Green's functions in (\ref{perteqnAd}-\ref{perteqnBd}) are 
identical to (\ref{Gret}). The free field expansions are similarly 
identical to (\ref{freefield}-\ref{ccr}), except that each field has 
an independent canonically normalized set of creation and annihilation 
operators:
\begin{eqnarray}
A_0(t,\vec{x}) & = &
\int \frac{d^3k}{(2\pi)^3} \;
\Biggl\{
e^{i \vec{k} \cdot \vec{x}} \, u(t,k) \; \alpha_{\vec{k}} \, + \,
e^{-i \vec{k} \cdot \vec{x}} \, u^*(t,k) \; \alpha^{\dagger}_{\vec{k}}
\, \Biggr\}
\;\; , \label{freefieldAd} \\
B_0(t,\vec{x}) & = &
\int \frac{d^3k}{(2\pi)^3} \;
\Biggl\{
e^{i \vec{k} \cdot \vec{x}} \, u(t,k) \; \beta_{\vec{k}} \, + \,
e^{-i \vec{k} \cdot \vec{x}} \, u^*(t,k) \; \beta^{\dagger}_{\vec{k}}
\, \Biggr\}
\;\; . \label{freefieldBd}
\end{eqnarray}

The second homogeneous term in (\ref{perteqnBd}) accounts for the
non-canonical commutation relation (\ref{noncan}). This second term 
can be expressed as the integral of a total derivative:
\begin{eqnarray}
&\!\!\!\! - \!\!\!\!& \frac{\lambda}2 \int d^3x' \; 
G_{\rm ret}\Bigl( t, \vec{x} \ ; 0, \vec{x}' \Bigr) 
\, A^2(0,\vec{x}') \, \dot{B}(0,\vec{x}') 
\nonumber \\
& \mbox{} &
= \, -\frac{\lambda}2 \int_0^t dt' \int d^3x' \,
\partial_{\mu}' \left\{ \,
\sqrt{-g(x')} \, g^{\mu\nu}(x') \, G_{\rm ret}\Bigl(x;x'\Bigr) \,
A^2(x') \, \partial_{\nu}' B(x') \, \right\} 
\;\; , \nonumber \\
& \mbox{} &
= \, -\frac{\lambda}2 \int_0^t dt' \, a^{\prime 3} 
\int d^3x' \; \partial_{\mu}' G_{\rm ret}\Bigl(x;x'\Bigr) \,
g^{\mu\nu}(x') \, A^2(x') \, \partial_{\nu}' B(x')
\nonumber \\
& \mbox{} & \hspace{1cm}  
- \, \lambda \int_0^t dt' \, a^{\prime 3} 
\int d^3x' \, G_{\rm ret}\Bigl(x;x'\Bigr) \left\{
\frac{ A(x') \, A_{,\mu}(x') \, B^{,\mu}(x')}
{1 + \frac{\lambda}2 A^2(x')} \right\} 
\;\; . \label{perteqnBdpart2}
\end{eqnarray}
In reaching this final expression we have made use of the equation of
motion (\ref{BEOM}). Combining terms gives our final form for the
Yang-Feldman equation of the field $B$,
\begin{eqnarray}
B(t,\vec{x}) & = &
B_0(t,\vec{x}) 
\label{BYF} \\
& \mbox{} &
- \, \frac{\lambda}2 \int_0^t dt' \, a^{\prime 3} \int d^3x' \; 
\partial_{\mu}' G_{\rm ret}\Bigl( t,\vec{x} \ ; t,\vec{x}' \Bigr)
\, g^{\mu\nu}(x') \, A^2(x') \, \partial_{\nu}' B(x') 
\;\; . \nonumber
\end{eqnarray}

$\bullet$ {\it Stochastic Realization:} We need to make the transition
from the full fields $A$ and $B$ to the stochastic fields 
$A_{\mbox{\tiny IR}}$ and $B_{\mbox{\tiny IR}}$. Most of this proceeds
as the reduction of Section 3, with slight modifications to accomodate
derivative interactions. The free fields suffer the same truncation of
their ultraviolet modes but we must now include the next order term in
the long wavelength expansion of the mode functions because the first
term has zero time derivative:
\begin{eqnarray}
A_0(t,\vec{x}) & \longrightarrow &
A_{\mbox{\tiny IR} \, 0}(t, \vec{x}) \; \equiv 
\label{irAd} \\
& \mbox{} & \hspace{-2cm}
\int \frac{d^3k}{(2\pi)^3} \;
\theta\Bigl( Ha(t) - k \Bigr) \,
\frac{H}{\sqrt{2 k^3}} \, 
\Biggl[ 1 + \frac12 \Bigl( \frac{k}{H a(t)} \Bigr)^2 \Biggr] \,
\Biggl\{ \,
e^{i \vec{k} \cdot \vec{x}} \, \alpha_{\vec{k}} \, + \,
e^{-i \vec{k} \cdot \vec{x}} \, \alpha^{\dagger}_{\vec{k}}
\, \Biggr\}
\;\; , \nonumber \\
B_0(t,\vec{x}) & \longrightarrow &
B_{\mbox{\tiny IR} \, 0}(t, \vec{x}) \; \equiv
\label{irBd} \\
& \mbox{} & \hspace{-2cm}
\int \frac{d^3k}{(2\pi)^3} \;
\theta\Bigl( Ha(t) - k \Bigr) \,
\frac{H}{\sqrt{2 k^3}} \, 
\Biggl[ 1 + \frac12 \Bigl( \frac{k}{H a(t)} \Bigr)^2 \Biggr] \, 
\Biggl\{ \,
e^{i \vec{k} \cdot \vec{x}} \, \beta_{\vec{k}} \, + \,
e^{-i \vec{k} \cdot \vec{x}} \, \beta^{\dagger}_{\vec{k}}
\, \Biggr\}
\;\; . \nonumber
\end{eqnarray}
The stochastic limit of the retarded Green's function is unchanged:
\begin{equation}
a^{\prime 3} \; G_{\rm ret}\Bigl( t,\vec{x} \ ; t',\vec{x}' \Bigr)
\quad \longrightarrow \quad
\frac1{3H} \, \theta(t-t') \; \delta^3(\vec{x} - \vec{x}')
\;\; . \label{irGretd}
\end{equation}

The new feature of these equations is that each interaction
contains two derivatives. From the analysis of Section 3 it is 
apparent that differentiating a propagator precludes it from
contributing an infrared logarithm. Said another way, it is always 
preferable to have a derivative act on the rapidly varying scale
factor rather than the slowly varying field. We, therefore, expect 
that the Hubble friction term dominates the scalar d'Alembertian:
\begin{equation}
B^{;\mu}_{~~\mu} \, = \,
-\ddot{B} - 3 H \dot{B} + \frac{\nabla^2}{a^2} B
\quad \longrightarrow  \quad
-3 H \dot{B}_{\mbox{\tiny IR}} 
\;\; . \label{KEruleBd}
\end{equation}
There are no d'Alembertians in our Yang-Feldman equations but the 
time derivative of the retarded Green's function in (\ref{BYF})
contributes a similar term. The space derivatives always give two
derivatives acting upon fields, and no term is comparably important 
to (\ref{KEruleBd}) when the two derivatives act on different fields:
\begin{eqnarray}
B_{,\mu} \; B^{,\mu} & = &
- {\dot B}^2 \, + \,
\frac{1}{a^2} {\vec \nabla}B \cdot {\vec \nabla}B
\quad \longrightarrow \quad 0
\;\; , \label{PEruleBBd} \\
A_{,\mu} \; B^{,\mu} & = &
- {\dot A}{\dot B} \, + \,
\frac{1}{a^2} {\vec \nabla}A \cdot {\vec \nabla}B
\quad \longrightarrow \quad 0
\;\; . \label{PEruleABd}
\end{eqnarray}
By dropping all double derivatives of the fields we arrive at
the equations which recover the stochastic fields:
\begin{eqnarray}
A_{\mbox{\tiny IR}}(t,\vec{x}) & = &
A_{\mbox{\tiny IR} \, 0}(t,\vec{x}) 
\;\; , \label{ireqnAd} \\
B_{\mbox{\tiny IR}}(t,\vec{x}) & = &
B_{\mbox{\tiny IR} \, 0}(t,\vec{x}) \, - \,
\frac{\lambda}2 \int_0^t dt' \; 
A^2_{\mbox{\tiny IR}}(t',\vec{x})
\dot{B}_{\mbox{\tiny IR}}(t',\vec{x}) 
\;\; . \qquad \label{ireqnBd} 
\end{eqnarray}

As in Section 3, the local stochastic field equations are obtained
by differentiating the stochastic Yang-Feldman equations 
(\ref{ireqnAd}-\ref{ireqnBd}) with respect to time. Also, as in Section 
3, derivatives of free fields produce stochastic source terms:
\begin{eqnarray}
\dot{A}_{\mbox{\tiny IR} \, 0}(t,\vec{x}) 
& \equiv &
f_{\mbox{\tiny A}}(t,\vec{x}) 
\label{stochsourceAd} \\
& = & 
\int \frac{d^3k}{(2\pi)^3} \;
\delta\Bigl( k \!-\! H a(t) \Bigr) \;
\frac{H^2}{\sqrt{2k}} \;
\Biggl\{ \, e^{i \vec{k} \cdot \vec{x}} \, \alpha_{\vec{k}} \, + \,
e^{-i \vec{k} \cdot \vec{x}} \, \alpha^{\dagger}_{\vec{k}} \, \Biggr\}
\nonumber \\
& \mbox{} & 
- \, \frac1{a^2(t)} \, \int \frac{d^3k}{(2\pi)^3} \;
\theta\Bigl( H a(t) \!-\! k \Bigr) \; \sqrt{\frac{k}{2}} \;
\Biggl\{ \, e^{i \vec{k} \cdot \vec{x}} \, \alpha_{\vec{k}} \, + \,
e^{-i \vec{k} \cdot \vec{x}} \, \alpha^{\dagger}_{\vec{k}} \, \Biggr\}
\;\; , \nonumber \\
\dot{B}_{\mbox{\tiny IR} \, 0}(t,\vec{x}) 
& \equiv &
f_{\mbox{\tiny B}}(t,\vec{x}) 
\label{stochsourceBd} \\
& = & 
\int \frac{d^3k}{(2\pi)^3} \;
\delta\Bigl( k \!-\! H a(t) \Bigr) \;
\frac{H^2}{\sqrt{2k}} \;
\Biggl\{ \, e^{i \vec{k} \cdot \vec{x}} \, \beta_{\vec{k}} \, + \,
e^{-i \vec{k} \cdot \vec{x}} \, \beta^{\dagger}_{\vec{k}} \, \Biggr\}
\nonumber \\
& \mbox{} & 
- \, \frac1{a^2(t)} \, \int \frac{d^3k}{(2\pi)^3} \;
\theta\Bigl( H a(t) \!-\! k \Bigr) \; \sqrt{\frac{k}{2}} \;
\Biggl\{ \, e^{i \vec{k} \cdot \vec{x}} \, \beta_{\vec{k}} \, + \,
e^{-i \vec{k} \cdot \vec{x}} \, \beta^{\dagger}_{\vec{k}} \, \Biggr\}
\;\; . \nonumber
\end{eqnarray}
Because no derivatives of $A{\mbox{\tiny IR}}$ appear in the Yang-Feldman 
stochastic field equations we can neglect the final term in
(\ref{stochsourceAd}) but it must be retained in (\ref{stochsourceBd}) on
account of the $\dot{B}_{\mbox{\tiny IR}}$ which appears in (\ref{ireqnBd}).
The resulting local stochastic equations are:
\begin{eqnarray}
\dot{A}_{\mbox{\tiny IR}}(t,\vec{x}) 
& = &
f_{\mbox{\tiny A}}(t,\vec{x}) 
\;\; , \label{stocheqnAd} \\
\dot{B}_{\mbox{\tiny IR}}(t,\vec{x}) 
& = &
f_{\mbox{\tiny B}}(t,\vec{x}) \, - \,
\frac{\lambda}2 \; 
A^2_{\mbox{\tiny IR}}(t,\vec{x}) \;
\dot{B}_{\mbox{\tiny IR}}(t,\vec{x}) 
\;\; . \qquad \label{stocheqnBd}
\end{eqnarray}
These are in good agreement with the equations which are obtained by 
directly applying the reduction rules I-II (\ref{ruleI}-\ref{ruleII}) 
of Section 3 to the full equations of motion (\ref{LdAeom}-\ref{LdBeom}):
\begin{eqnarray}
-3 H \Bigl( \dot{A}_{\mbox{\tiny IR}} - f_{\mbox{\tiny A}} \Bigr)
& = & 0 \;\; , \label{stocheqnAd2} \\
-3 H \Bigl( \dot{B}_{\mbox{\tiny IR}} - f_{\mbox{\tiny B}} \Bigr)
\; - \; \frac{3 H \lambda}2 \, A^2_{\mbox{\tiny IR}} \,
\dot{B}_{\mbox{\tiny IR}}
& = & 0 \;\; . \label{stocheqnBd2} 
\end{eqnarray}
Note the curious fact that interactions containing derivatives are 
{\it free} of stochastic source terms.
 
It is amusing to note that we can obtain explicit operator solutions
to the stochastic field equations (\ref{stocheqnAd}-\ref{stocheqnBd}). 
The exact solution for $A_{\mbox{\tiny IR}}$ follows trivially from 
(\ref{stocheqnAd}):
\begin{equation}
A_{\mbox{\tiny IR}}(t, \vec{x}) \; = \;
\int_0^t dt' \; f_{\mbox{\tiny A}}(t', \vec{x}) \; = \;
A_{\mbox{\tiny IR} \, 0}(t, \vec{x})
\;\; , \label{stochsolAd} \\
\end{equation}
and shows that $A_{\mbox{\tiny IR}}$ receives no corrections to its
free field value $A_{\mbox{\tiny IR} \, 0}$. A few simple re-arrangements
in (\ref{stocheqnBd}) give the closed form operator
solution for $B_{\mbox{\tiny IR}}$:
\begin{equation}
B_{\mbox{\tiny IR}}(t, \vec{x})
\, = \,
\int_0^t dt' \;
\frac{f_{\mbox{\tiny B}}(t', \vec{x})}
{1 \, + \, \frac12 \lambda \,
A^2_{\mbox{\tiny IR} \, 0}(t', \vec{x})}
\;\; . \label{stochsolBd}
\end{equation}
From (\ref{stochsolBd}) we conclude that $B_{\mbox{\tiny IR}}$ exhibits
a kind of spacetime-dependent field strength renormalization whereby
each set of $\beta$-modes which experiences horizon crossing is attenuated
by the factor $1 + \frac{\lambda}{2} A^2_{\mbox{\tiny IR} \, 0}$. In other
words, the stochastic field $B_{\mbox{\tiny IR}}$ acquires corrections
from its free field form which diminish with time and it reaches some
constant value asymptotically. \\

\begin{figure}
\centerline{\epsfig{file=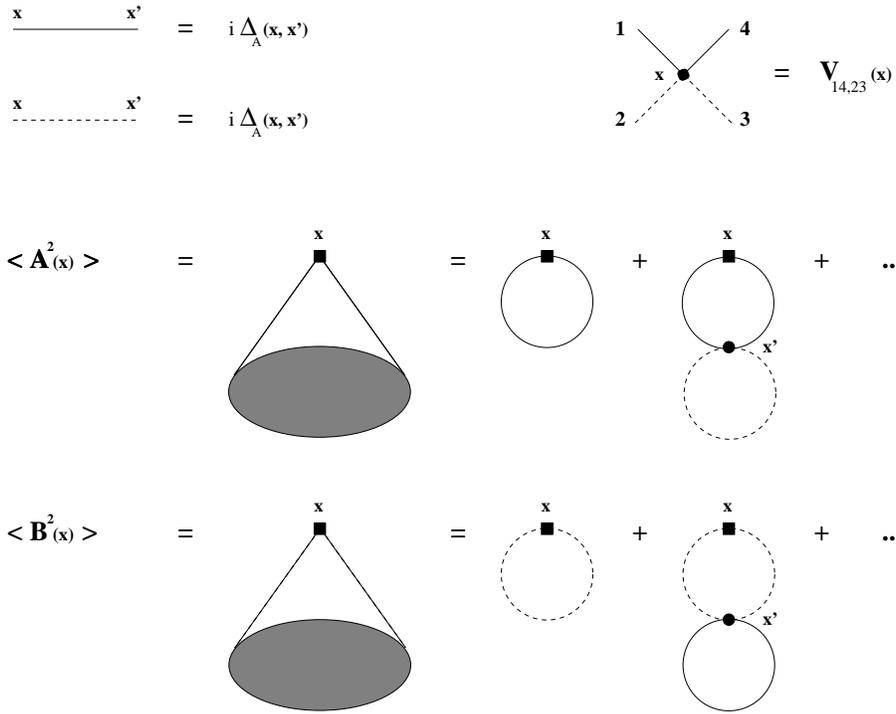,height=3.7in}}
\caption{\footnotesize The Feynman rules and the diagrammatic expansion
to $O(\lambda)$ of the
\break \mbox{} \hspace{1.8cm}
vacuum expectation value of the squares of the fields in the scalar model
with derivative interactions. Solid lines correspond to the scalar field
$A$ and segmented lines to the scalar field $B$.}
\end{figure}

$\bullet$ {\it Perturbative Agreement:} It is possible to calculate the 
one and two loop contributions to the expectation values of the squares of
the full fields $A$ and $B$. Because they are both massless and minimally 
coupled fields, they have the same propagator given by (\ref{DeltaA}). 
The basic vertex is:
\begin{equation}
V_{14,23}(x) \; = \;
-i\lambda \, \sqrt{-g} \, g^{\mu\nu} \,
\partial_{1 \mu} \; \partial_{4 \nu}
\;\; , \label{dvertex}
\end{equation}
where we have used the notation of Figure 1 which also depicts all
relevant graphs. In the Appendix we use these Feynman rules to compute:
\begin{eqnarray}
\left\langle \,
A^2 (t, {\vec x})
\, \right\rangle_{\mbox{\tiny IR}}
&\!\! = \!\!&
\left( \frac{H}{2\pi} \right)^2 \ln a
\, + \, O(\lambda^2)
\;\; , \label{pertAd2pt} \\
\left\langle \,
B^2 (t, {\vec x})
\, \right\rangle_{\mbox{\tiny IR}}
&\!\! = \!\!&
\left( \frac{H}{2\pi} \right)^2 \ln a
\, + \,
\frac{\lambda H^4}{2^6 \pi^4}
\Big[ -\ln^2 a \, + \, O(\ln a) \, \Big]
\, + \, O(\lambda^2)
\;\; . \qquad \label{pertBd2pt}
\end{eqnarray}

It is apparent that (\ref{pertAd2pt}) agrees with the stochastic result 
(\ref{stochsolAd}). To see that (\ref{pertBd2pt}) agrees as well with
(\ref{stochsolBd}), we perturbatively expand the stochastic operator
solution:
\begin{equation}
B_{\mbox{\tiny IR}}(t, \vec{x}) \; = \;
B_{\mbox{\tiny IR} \, 0}(t, \vec{x}) \, - \,
\frac{\lambda}2 \int_0^t dt' \,
A^2_{\mbox{\tiny IR}}(t',\vec{x}) \; 
f_{\mbox{\tiny B}}(t',\vec{x})
\, + \, O(\lambda^2) 
\;\; , \label{pertBd1}
\end{equation}
and square it:
\begin{equation}
B^2_{\mbox{\tiny IR}}(t, \vec{x}) \; = \;
B^2_{\mbox{\tiny IR} \, 0}(t, \vec{x}) \, - \, 
\lambda \, B_{\mbox{\tiny IR}}(t, \vec{x}) 
\int_0^t dt' \,
A^2_{\mbox{\tiny IR}}(t',\vec{x}) \; 
f_{\mbox{\tiny B}}(t',\vec{x})
\, + \, O(\lambda^2) 
\;\; . \qquad \label{pertBd2}
\end{equation}
Taking the vacuum expectation value of (\ref{pertBd2}) gives:
\begin{eqnarray}
\lefteqn{\Big\langle \; 
B^2_{\mbox{\tiny IR}}(t, \vec{x}) 
\; \Big\rangle_{\mbox{\tiny VEV}} 
\; = \;
\Big\langle \; B^2_{\mbox{\tiny IR} \, 0}(t, \vec{x}) 
\; \Big\rangle_{\mbox{\tiny VEV}} } 
\label{pertBd2vev1} \\
& & - \, \lambda \int_0^t dt' \;
\Big\langle \; A^2_{\mbox{\tiny IR} \, 0}(t',\vec{x})
\; \Big\rangle_{\mbox{\tiny VEV}} \;
\Big\langle \; B_{\mbox{\tiny IR} \, 0}(t,\vec{x}) \; 
f_{\mbox{\tiny B}}(t',\vec{x})
\; \Big\rangle_{\mbox{\tiny VEV}}
+ \, O(\lambda^2)
\;\; . \nonumber
\end{eqnarray}
For the case in which $t > t'$ we have:
\begin{eqnarray}
\Big\langle \; B_{\mbox{\tiny IR} \, 0}(t,\vec{x}) \;
f_{\mbox{\tiny B}}(t',\vec{x}) \; \Big\rangle_{\mbox{\tiny VEV}} 
& = & 
\frac{H^2}{4 \pi^2} \int \frac{dk}{k} \,
\Biggl\{ H^2 a' \; \delta(Ha' - k) 
\label{freeBd2ptvev} \\
& \mbox{} & \hspace{2cm}
- \, \frac{k^2}{H a^{\prime \, 2}} \, \theta(Ha' - k) \Biggr\} 
\; = \; \frac{H^3}{8\pi^2}
\;\; . \nonumber
\end{eqnarray}
Note that we discard the time dependent corrections to the wave function
in $B_{\mbox{\tiny IR}}$ which would in any case give sub-dominant terms 
like $a^{-2} a^{\prime \, 2}$. Hence we obtain:
\begin{eqnarray}
\Big\langle \; B^2_{\mbox{\tiny IR}}(t, \vec{x}) 
\; \Big\rangle_{\mbox{\tiny VEV}} 
& = & 
\frac{H^2}{4 \pi^2} \, \ln a \, - \,
\lambda \int_0^t dt' \; 
\frac{H^2}{4 \pi^2} \; \ln a' \; \frac{H^3}{8 \pi^2} 
\; + \; O(\lambda^2) 
\;\; , \qquad \label{pertBd2vev2} \\
& = & 
\frac{H^2}{4 \pi^2} \, \ln a  \, - \,
\frac{\lambda H^4}{2^6 \pi^4} \, \ln^2 a 
\; + \; O(\lambda^2) 
\;\; , \label{pertBd2vev3}
\end{eqnarray}
which establishes the non-trivial agreement between (\ref{stochsolBd})
and (\ref{pertBd2pt}).
 
\section{Constrained Fields}

Besides dynamical degrees of freedom, gravitation contains constrained
fields. Because these fields possess no dynamical degrees of freedom
there should be no independent stochastic sources for them. However, a 
constrained variable can depend non-linearly and non-locally upon the 
dynamical fields and it is by no means clear how to include the extra 
stochastic jitter the constrained field acquires due to each increment 
of dynamical modes which experience horizon crossing.

No purely scalar model can mimic the gravitational system we really 
need to understand. It would be interesting -- and it seems feasible 
-- to gain insight into constrained degrees of freedom by attempting 
a stochastic formulation of scalar quantum electrodynamics. This model 
certainly possesses a gauge constraint, and the one loop vacuum 
polarization has been shown to contain an infrared logarithm \cite{PTW}. 
However, as a first step we shall attempt to gain insight from purely 
scalar models by exploiting the existence of a covariant gauge for 
gravity in which all components of the perturbed metric are either 
minimally coupled or else conformally coupled \cite{AW}.

The model we shall study consists of a massless, minimally coupled 
scalar $A$, a massless, conformally coupled scalar $C$ and an elementary
interaction between the two:
\begin{eqnarray}
{\cal L}_{\mbox{\tiny C}} & = &
- \frac12 \, \sqrt{-g} \, g^{\mu\nu} \,
\partial_{\mu} A \; \partial_{\nu} A \;
- \frac12 \, \sqrt{-g} \, g^{\mu\nu} \,
\partial_{\mu} C \; \partial_{\nu} C
\nonumber \\
& \mbox{} &
- \frac{D-2}{8(D-1)} \, \sqrt{-g} \, R \; C^2 \;
- \frac12 \, \kappa H^2 \, \sqrt{-g} \; A^2 \; C
\;\; . \label{Lc}
\end{eqnarray}
The equations of motion are:
\begin{eqnarray}
\frac1{\sqrt{-g}} \, \frac{\delta S}{\delta A}
& = &
A^{;\mu}_{~~\mu} \, - \, \kappa H^2 A \, C
\; = \; 0
\;\; , \label{LcAeom} \\
\frac1{\sqrt{-g}} \, \frac{\delta S}{\delta C}
& = &
C^{;\mu}_{~~\mu} \, - \, \frac16 R \, C \, - \,
\frac12 \kappa H^2 A^2
\nonumber \\
& = &
C^{;\mu}_{~~\mu} \, - \, 2H^2 \, C \, - \,
\frac12 \kappa H^2 A^2
\; = \; 0
\;\; . \label{LcCeom}
\end{eqnarray}
As usual, the perturbative initial value solution comes from iterating
the Yang-Feldman equations:
\footnote{Field arguments are sometimes compressed to 4-vector notation:
$(t, {\vec x}) \equiv x$.}
\begin{eqnarray}
A(t,\vec{x}) &\!\!\! = \!\!\!&
A_0(t,\vec{x}) -
\kappa H^2 \!\!\int_0^t dt' \, a^3(t')
\int d^3x' \, G_{\rm ret} \Bigl( t,\vec{x} \ ; t',\vec{x}' \Bigr)
\, A(x') \, C (x')
\hspace{1.1cm} \label{perteqnAc} \\
C(t,\vec{x}) &\!\!\! = \!\!\!&
C_0(t,\vec{x}) -
\frac12 \kappa H^2 \!\!\int_0^t dt' \, a^3(t')
\int d^3x' \, G_{\rm ret}^C\Bigl( t,\vec{x} \ ; t',\vec{x}' \Bigr)
\; A^2 (x')
\; . \label{perteqnCc}
\end{eqnarray}
The minimally coupled free field $A_0$ is identical to (\ref{freefieldAd}). 
Its conformally coupled cousin is:
\begin{equation}
C_0(t,\vec{x}) \, = \,
\int \frac{d^3k}{(2\pi)^3} \;
\Biggl\{
e^{i \vec{k} \cdot \vec{x}} \, v(t,k) \; \gamma_{\vec{k}} \, + \,
e^{-i \vec{k} \cdot \vec{x}} \, v^*(t,k) \; \gamma^{\dagger}_{\vec{k}}
\, \Biggr\}
\;\; , \label{freefieldCc}
\end{equation}
where $\gamma^{\dagger}_{\vec{k}}$ and $\gamma_{\vec{k}}$ are canonically 
normalized creation and annihilation operators and $v(t,k)$ is the 
conformal mode function:
\begin{equation}
v(t,k) \; = \;
\frac{-i}{\sqrt{2 k}} \, \frac{\exp\Bigl[\frac{ik}{Ha(t)}\Bigr]}{a(t)}
\;\; . \label{modefnctsCc}
\end{equation}
The minimally coupled retarded Green's function $G_{\rm ret}(t,\vec{x};
t',\vec{x}')$ is the same one (\ref{Gret}) we have been using since Section 
3. The conformally coupled retarded Green's function is:
\begin{equation}
G_{\rm ret}^C\Bigl( t,\vec{x} \ ; t',\vec{x}' \Bigr) \, = \,
\frac{H^2}{4\pi} \; \theta(t - t') \,
\Bigg\{ \, \frac{\delta\Bigl( H \Vert\vec{x} - \vec{x}'\Vert +
\frac1{a} - \frac1{a'} \Bigr)} {a a' \, H \Vert\vec{x} - \vec{x}'\Vert}
\, \Bigg\}
\;\; . \label{GretCc}
\end{equation}

$\bullet$ {\it Stochastic Realization:} We seek a simplification of the 
Yang-Feldman equations for stochastic fields $A_{\mbox{\tiny IR}}(t,\vec{x})$ 
and $C_{\mbox{\tiny IR}}(t,\vec{x})$ which preserves the leading infrared 
logarithms. The $A$ equation (\ref{perteqnAc}) can be reduced as in the 
previous examples:
\begin{equation}
A_{\mbox{\tiny IR}}(t,\vec{x}) \; = \;
A_{\mbox{\tiny IR} \, 0}(t,\vec{x}) \, - \,
\frac{\kappa H}3 \int_0^t dt' \, 
A_{\mbox{\tiny IR}}(t',\vec{x}) \,
C_{\mbox{\tiny IR}}(t',\vec{x}) 
\;\; . \label{eqnAcIR}
\end{equation}
The free stochastic mode sum $A_{\mbox{\tiny IR} \, 0}(t,\vec{x})$ 
-- as in the previous examples -- is given by (\ref{freefieldAd}). 
We obtain the local stochastic field equation by differentiation:
\begin{equation}
{\it Leading \; Infrared}
\quad \Longrightarrow \quad
\dot{A}_{\mbox{\tiny IR}}(t,\vec{x}) \; = \;
f_{\mbox{\tiny A}}(t,\vec{x}) \, - \, 
\frac{\kappa H}3 \,
A_{\mbox{\tiny IR}}(t,\vec{x}) \,
C_{\mbox{\tiny IR}}(t,\vec{x}) 
\qquad
\label{eqnAcLIR}
\end{equation}
The stochastic source $f_{\mbox{\tiny A}}(t,\vec{x})$ is given by  
(\ref{stochsourceAd}).

To obtain the analogous reduction for the $C$ equation (\ref{perteqnCc})
note first that a free field does not produce infrared logarithms unless
its mode function goes like $k^{-\frac32}$ near $k=0$. The conformally
coupled mode function (\ref{modefnctsCc}) goes like $k^{-\frac12}$ so 
its associated free field truncates to zero:
\begin{equation}
{\it Leading \; Infrared}
\quad \Longrightarrow \quad 
C_0(t,\vec{x}) \, \longrightarrow \, 
C_{\mbox{\tiny IR} \, 0}(t,\vec{x}) = 0
\;\; . \label{freefieldCcLIR}
\end{equation}
Then, we determine the infrared limit of the conformally coupled Green's 
function by beginning with the generic expression:
\begin{eqnarray}
G_{\rm ret}^C \Bigl( t,\vec{x} \ ; t',\vec{x}' \Bigr)
& = &
i \theta(t-t') \times
\label{fullGretC} \\
& &
\int \frac{d^3k}{(2\pi)^3} \,
\Bigl[ \, v(t,k) \; v^*(t',k) - v^*(t,k) \; v(t',k) \, \Bigr] \,
e^{i \vec{k} \cdot \Delta \vec{x}}
\;\; . \quad \nonumber
\end{eqnarray}
Using (\ref{modefnctsCc}) the long wavelength expansion of the 
bracketed term is:
\begin{equation}
\Bigl[ \, v(t,k) \; v^*(t',k) - v^*(t,k) \; v(t',k) \, \Bigr] \, = \,
\frac{i}{H} \Biggl(
\frac1{a a^{\prime 2}} - \frac1{a^2 a^{\prime}} \Biggr)
\, + \, O(k^2)
\;\; . \label{vv*exp}
\end{equation}
Consequently, the infrared limit of the conformally coupled Green's
function is:
\begin{equation}
a^{\prime 3} \; G_{\rm ret}^{C \hspace{0.05cm} \mbox{\tiny IR}}
\Bigl( t,\vec{x} \ ; t',\vec{x}' \Bigr)
\; \longrightarrow \;
\frac1{H} \, \theta(t-t') \; \delta^3(\vec{x} - \vec{x}') \,
\Biggl( \frac{a'}{a} - \frac{a^{\prime 2}}{a^2} \Biggr)
\;\; . \label{irGretC}
\end{equation}
The stochastic Yang-Feldman equation for $C_{\mbox{\tiny IR}}$ could 
be written as:
\begin{equation}
C_{\mbox{\tiny IR}} (t,\vec{x}) \; = \;
- \frac{\kappa H}{2} \int_0^t dt' \;
\Biggl( \frac{a^{\prime}}{a} - \frac{a^{\prime 2}}{a^2} \Biggr)
\; A^2_{\mbox{\tiny IR}} (t',\vec{x}')
\; . \label{perteqnCcLIR}
\end{equation}
However, a further simplification is possible. By neglecting the slow 
variation in $A^2_{\mbox{\tiny IR}}(t'\vec{x})$ relative to the rapidly 
growing scale factors we can perform the intergation over $t'$:
\begin{equation}
{\it Leading \; Infrared}
\quad \Longrightarrow \quad
C_{\mbox{\tiny IR}} (t,\vec{x}) \; = \;
- \frac{\kappa}4 \; A^2_{\mbox{\tiny IR}} (t,\vec{x})
\; . \label{eqnCcLIR}
\end{equation}
Substituting (\ref{eqnCcLIR}) for $C_{\mbox{\tiny IR}}$ in 
(\ref{eqnAcLIR}) reveals the amusing fact that $A_{\mbox{\tiny IR}}$ 
obeys the same stochastic field equation as the self-interacting scalar 
$\phi$ of Section 3 with the identification $\lambda = -\frac32 
\kappa^2 H^2$. Because the effective 4-point coupling is negative this 
system ought to exhibit a runaway instability.

Equations (\ref{eqnAcLIR}) and (\ref{eqnCcLIR}) result from applying the 
reduction rules I-II (\ref{ruleI}-\ref{ruleII}) of Section 3 to the full 
equations of motion (\ref{LcAeom}-\ref{LcCeom}). First apply rules I-II
(\ref{ruleI}-\ref{ruleII}) to the kinetic term of the minimally coupled 
scalar in the usual way:
\begin{equation}
A^{;\mu}_{~~\mu} \; = \;
- {\ddot A} \, - \, 3 H {\dot A} \, + \,
\frac{\nabla^2}{a^2} \, A
\quad \longrightarrow \quad
- 3 H \Bigl( {\dot A}_{\mbox{\tiny IR}} - f_{\mbox{\tiny A}} \Bigr)
\;\; . \label{KEruleAc}
\end{equation}
By rule I (\ref{ruleI}) the leading contribution to the conformally 
coupled kinetic term is undifferentiated, so by rule II (\ref{ruleII})
there is no stochastic source:
\begin{equation}
C^{;\mu}_{~~\mu} \, - \, 2 H^2 C \; = \;
- {\ddot C} \, - \, 3 H {\dot C} \, + \,
\frac{\nabla^2}{a^2} \, C \, - \, 2 H^2 C
\quad \longrightarrow \quad
-2 H^2 C_{\mbox{\tiny IR}}
\;\; . \label{KEruleCc}
\end{equation}
The complete stochastic field equations are therefore:
\begin{eqnarray}
-3H \Bigl( \dot{A}_{\mbox{\tiny IR}} - f_{\mbox{\tiny A}} \Bigr)
\, - \, \kappa H^2 \, A_{\mbox{\tiny IR}} \; C_{\mbox{\tiny IR}}
& = & 0
\;\; , \label{stocheqnAc} \\
-2H^2 \, C_{\mbox{\tiny IR}} \, - \,
\frac12 \kappa H^2 \, A_{\mbox{\tiny IR}} \; C_{\mbox{\tiny IR}}
& = & 0
\;\; , \label{stocheqnCc}
\end{eqnarray}
and the rules I-II (\ref{ruleI}-\ref{ruleII}) are again shown to
give correct results. \\

\begin{figure}
\centerline{\epsfig{file=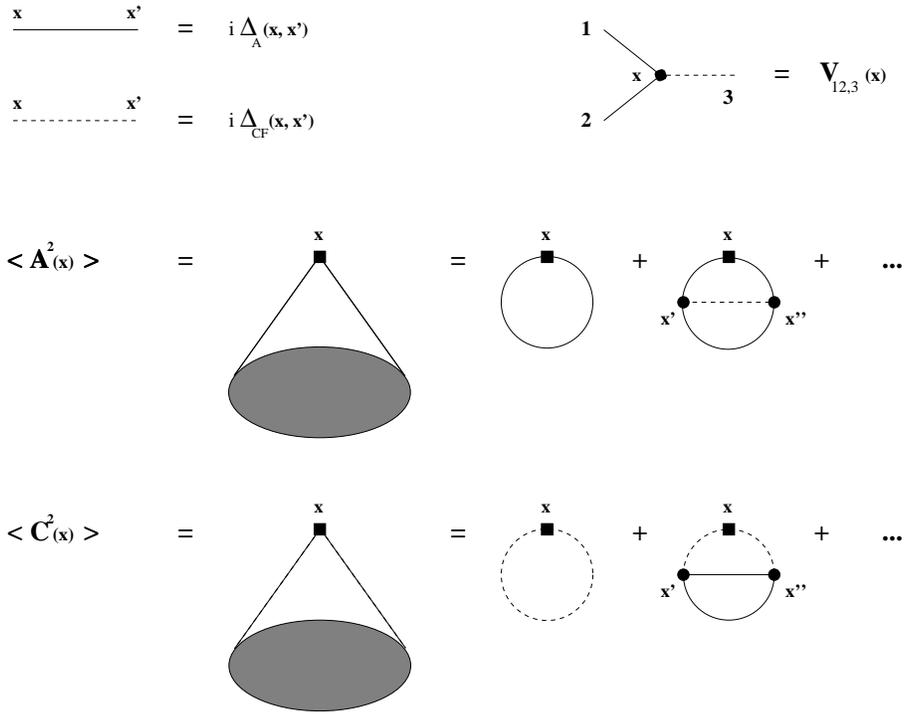,height=3.7in}}
\caption{\footnotesize The Feynman rules and the diagrammatic expansion
to $O(\kappa^2)$ of the
\break \mbox{} \hspace{1.8cm}
vacuum expectation value of the squares of the fields in the scalar model
with constraints. Solid lines correspond to the scalar field $A$ and
segmented lines to the scalar field $C$.}
\end{figure}

$\bullet$ {\it Perturbative Agreement:} It is simple to compute the 
expectation values of squares of the full quantum fields $A$ and $C$ 
to a few orders in perturbation theory. The Feynman rules and 
relevant diagrams are shown in Figure 2; the explicit form of the basic 
vertex is:
\begin{equation}
V_{12,3}(x) \; = \;
-i\kappa \, H^2
\;\; . \label{cvertex}
\end{equation}
The lowest order contributions are just the coincidence limits of
the two propagators:
\begin{eqnarray}
\left\langle \;
A^2 (t, {\vec x})
\; \right\rangle_{\mbox{\tiny VEV}}
& = & 
i\Delta_A(x;x) + O(\kappa^2) 
\; = \;
\frac{H^2}{4\pi^2} 
\Bigl\{ {\rm UV} + \ln a\Bigr\} 
+ O(\kappa^2)
\; , \qquad \label{pertAc2pt} \\
\left\langle \;
C^2 (t, {\vec x})
\; \right\rangle_{\mbox{\tiny VEV}}
& = & 
i\Delta{\mbox{\tiny CF}}(x;x) + O(\kappa^2) 
\; = \;
0 + O(\kappa^2)
\; , \label{pertCc2pt}
\end{eqnarray}
where ``UV'' stands for a constant, ultraviolet divergence. These
results are completely consistent with the perturbative solution 
of the stochastic equations (\ref{eqnAcLIR}) and (\ref{eqnCcLIR}):
\begin{eqnarray}
A_{\mbox{\tiny IR}}(t, \vec{x})
& = &
A_{\mbox{\tiny IR} \, 0}(t, \vec{x}) \; + \; 
\frac{\kappa^2 H}{12}
\int_0^t dt' \; A^3_{\mbox{\tiny IR} \, 0}(t',\vec{x}) 
\; + \; O(\kappa^4)
\;\; , \label{pertstochsolAc} \\
C_{\mbox{\tiny IR}}(t, {\vec x})
& = & 
-\frac{\kappa}4 \, A^2_{\mbox{\tiny IR} \, 0}(t,\vec{x}) 
\; + \; O(\kappa^3)
\;\; . \label{pertstochsolCc}
\end{eqnarray}
Based on the above expansions (\ref{pertstochsolAc}-\ref{pertstochsolCc})
we make the following predictions for the next order -- that is, two loop 
-- perturbative results:
\begin{eqnarray}
\left\langle \;
A^2_{\mbox{\tiny IR}}(t, {\vec x}) 
\; \right\rangle_{\mbox{\tiny VEV}}
& = & 
\frac{H^2}{4\pi^2} \ln a \; + \;
\frac{\kappa^2 H^4}{2^5 \, 3 \pi^4} \ln^3 a \; + \;
O(\kappa^4) 
\;\; , \label{LcpertA2ptpred} \\
\left\langle \; 
C^2_{\mbox{\tiny IR}}(t, {\vec x}) 
\; \right\rangle_{\mbox{\tiny VEV}}
& = & 
\frac{3 \kappa^2 H^4}{2^8 \pi^4} \ln^2 a \; + \; O(\kappa^4) 
\;\; . \label{LcpertC2ptpred} \\
\end{eqnarray}
It will be interesting to check these predictions.

\section{Epilogue}

The subject of this paper has been infrared logarithms. These
are powers of the number of inflationary e-foldings from when the 
system was released in a prepared state. They can arise in the 
expectation values of operators in quantum field theories which 
contain massless, minimally coupled scalars or gravitons. We
demonstrated that infrared logarithms derive entirely from long 
wavelength quanta. In Section 3 we were able to obtain a 
completely finite system, with a conserved stress tensor, by
truncating the ultraviolet modes from the free field of the 
Yang-Feldman equation. 

Infrared logarithms are very exciting because they may signal
important quantum effects. However, their continued growth
implies the breakdown of perturbation theory. A natural approach
for obtaining non-perturbative information is the leading 
logarithm approximation in which one attempts to sum the series
comprised of just the leading infrared logarithms at each order.
Starobinski\u{\i} has long argued that his formalism of stochastic
inflation \cite{starobinsky2,starobinsky3} recovers the leading
infrared logarithms for scalar fields with non-derivative interactions.
We have confirmed this using the infrared-truncated Yang-Feldman 
equation of Section 3. The leading infrared logarithms are not changed 
when the free field mode functions are simplified to retain only the 
first nonzero term in the long wavelength expansion and a similar 
simplification is done to the retarded Green's function. The fields
of the resulting system are commuting, even though they depend upon
creation and annihilation operators, and they obey precisely 
Starobinski\u{\i}'s Langevin equation.

Whereas there is little hope of obtaining non-perturbative solutions
for interacting quantum field theories in $3+1$ dimensions, classical
field equations {\it can} sometimes be solved exactly. And {\it simple}
classical equations are especially likely to possess exact solutions.
In Section 4 we reviewed the exact solution obtained by Starobinski\u{\i}
and Yokoyama \cite{starobinsky3} for the late time limit of a massless, 
minimally coupled scalar with a quartic self-interaction. We also reviewed 
Starobinski\u{\i}'s technique \cite{starobinsky5,FMSVV} for predicting
the infrared logarithms of explicit perturbative computations, and we
worked out the general result at orders $\lambda^0$, $\lambda^1$ and
$\lambda^2$.

The remainder of the paper concerned extending Starobinski\u{\i}'s
formalism to more complicated theories with derivative interactions
and constraints. We have written down a set of simple rules 
(\ref{ruleI}-\ref{ruleII}) for accomplishing this. By going through
the same procedure of first infrared-truncating the Yang-Feldman equations
and then making the long wavelength approximation on their mode functions
and retarded Green's function, we showed that rules I-II give the correct
results for a scalar model with derivative interactions (Section 5) and
for a reasonable scalar analogue of a model with gauge constraints
(Section 6). We also checked our rules against explicit perturbative
computations. 

We emphasize that just because the rules I-II (\ref{ruleI}-\ref{ruleII}) 
can be stated generally does not mean they apply generally. For any
particular model the stochastic Langevin equations implied by our
rules should be derived by infrared-truncating the Yang-Feldman
equations and then imposing the long wavelength approximation. 
Specific perturbative results should also be checked. 

The primary motivation for developing these rules is their application
to gravitation where some interesting perturbative results already
exist \cite{nctrpw1,nctrpw2}. We close with an amusing and thought-provoking
argument which is completely independent of these results and assumes
only that a stochastic formulation of inflationary quantum gravity exists.
If so, then it is straightforward to prove that back-reaction {\it must}
become important. Suppose the null hypothesis is correct: there is no 
significant back-reaction at any order. In that case the stochastic metric 
field is just the inflationary background plus the free field mode sum of 
long wavelength gravitons in the transverse-traceless field 
$h_{ij}^{\mbox{\tiny TT}}(t, {\vec x})$:
\begin{equation}
Null \; Hypothesis
\quad \Longrightarrow \quad
g_{ij}(t, {\vec x}) \; = \;
a^2(t) \left[ \,  \delta_{ij} \, + \,
h_{ij}^{\mbox{\tiny TT}}(t,{\vec x})
\, \right] \;\; . \label{nullhyp}
\end{equation}
Hence, the volume element operator is:
\begin{equation}
\sqrt{-g(t, {\vec x})} \; = \;
a^3(t) \left[ \, 1  \, - \,
\frac14 h_{ij}^{\mbox{\tiny TT}}(t, {\vec x}) \,
h_{ij}^{\mbox{\tiny TT}}(t, {\vec x}) 
\, + \, O \left( h^3 \right) \, \right]
\;\; . \label{dV}
\end{equation}
Its stochastic vacuum expectation value is simple to compute:
\begin{equation}
\left\langle \;
\sqrt{-g(t, {\vec x})}
\,\, \right\rangle_{\mbox{\tiny VEV}}
\; = \;
a^3(t) \left[ \, 1  \, - \,
\frac{4}{\pi} \, G H^2 \ln a \, + \,
O(G^2) \, \right]
\;\; . \label{volumeVEV}
\end{equation}
For large observation times this eventually shrinks to zero, at which
point back-reaction must have become significant. Hence it cannot be
consistent to ignore back-reaction!

In addition to illustrating the power of the stochastic formalism, the
argument we have just given establishes the need to include quantum
gravitational back-reaction. However, this does not necessarily mean
there will be significant modifications to the de Sitter background.
It may be instead that non-linear corrections to the graviton field,
and to the various constrained fields, add up to give only a small
shift that approaches a constant at late times. Precisely this happens
to the scalar model considered in Section 4. However, it was obvious
in this model that the classical force pushing the scalar down its 
potential would eventually compensate the tendency of inflationary
particle production to push it ever higher. No such mechanism is
apparent in gravity. Indeed, the physical picture seems rather to be
that the self-gravitation from continual production of inflationary
gravitons must eventually bring the universe to the verge of 
gravitational collapse. It would be as premature to ignore the
possibility of a significant effect from quantum gravity on the basis
of the scalar model as it would have been to discount the prospects for
asymptotic freedom -- in a somewhat different leading logarithm 
approximation --  based on the positive beta functions of QED and 
$\phi^4$ theory.


\newpage

\centerline{\bf Acknowledgements}

We are grateful to Alexei Starobinski\u{\i} for explaining his 
technique and for emphasizing that it recovers the leading infrared
logarithms of inflationary quantum field theory. We also thank Pei-Ming
Ho for finding a factor of two mistake in our initial treatment of
the model of Section 5, and Igor Volovich for discussions concerning
the stochastic limit of quantum field theory. This work was partially
supported by the European Social fund and National resources
Y$\Pi$E$\Pi\Theta$-PythagorasII-2103, by European Union grants FP-6-012679
and MRTN-CT-2004-512194, by NSF grant PHY-0244714, and by the Institute
for Fundamental Theory at the University of Florida.

\newpage

\section{Appendix: Perturbative Computations}

We shall derive expression (\ref{pertBd2pt}) from Section 5. The one
loop contribution derives from the coincidence limit (\ref{propAcoinc}
of the $A$-type propagator. At two loop order the contribution is:
\begin{eqnarray}
\left\langle \,
B^2 (t, {\vec x})
\, \right\rangle^{\mbox{\tiny 2-loop}}_{\mbox{\tiny VEV}}
& = &
\int d^Dx' \;
\sqrt{-g(x')} \; g^{\mu\nu}(x')
\biggl[ \, -\frac{i\lambda}{2} \; {i\Delta}_{A}(x';x')
- i \, \delta Z \, \biggr] \times \quad
\nonumber \\
& \mbox{} & \hspace{1cm}
\biggl\{ 
\left[ \partial_{\mu}^{\prime} \, {i\Delta}_{A ++}(x;x') \right]
\partial_{\nu}^{\prime} \, {i\Delta}_{A ++}(x;x')
\nonumber \\
& \mbox{} & \hspace{1.5cm}
- \left[ \partial_{\mu}^{\prime} \, {i\Delta}_{A +-}(x;x') \right]
\partial_{\nu}^{\prime} \, {i\Delta}_{A +-}(x;x') \biggr\}
\;\; , \label{app1}
\end{eqnarray}
where we have used the Feynman rules of Figure 1 and equations
(\ref{DeltaA}), (\ref{dvertex}). In (\ref{app1}), the subscript pairs
$++$ and $+-$ refer to the two possible variations of the interaction
vertex at $x'$ as required when computing true expectation values in
quantum field theory \cite{schwinger,jordan,nctrpw2}. The symbol
$\delta Z$ represents the field strength renormalization necessary to 
absorb the ultraviolet divergence coming from the coincidence limit
(\ref{propAcoinc}) of the propagator at the interaction point $x'$.

We first partially integrate both the $++$ and the $+-$ sectors of
(\ref{app1}) using the identity:
\begin{eqnarray}
\sqrt{-g(x')} \; g^{\mu\nu}(x')
\left[ \partial_{\mu}^{\prime} \, {i\Delta}_{A}(x;x') \right]
\partial_{\nu}^{\prime} \, {i\Delta}_{A}(x;x')
& = &
\nonumber \\
& \mbox{} & \hspace{-7cm}
\partial_{\mu}^{\prime} \left\{
\sqrt{-g(x')} \; g^{\mu\nu}(x') \;
{i\Delta}_{A}(x;x')
\partial_{\nu}^{\prime} \, {i\Delta}_{A}(x;x')
\right\}
\nonumber \\
& \mbox{} & \hspace{-7cm}
- \, {i\Delta}_{A}(x;x') \;
\partial_{\mu}^{\prime} \left\{
\sqrt{-g(x')} \; g^{\mu\nu}(x')
\partial_{\nu}^{\prime} \, {i\Delta}_{A}(x;x')
\right\}
\;\; . \qquad \label{app2}
\end{eqnarray}
Because the propagators obey:
\begin{eqnarray}
\partial_{\mu}^{\prime} \left\{
\sqrt{-g(x')} \; g^{\mu\nu}(x')
\partial_{\nu}^{\prime} \, {i\Delta}_{A ++}(x;x')
\right\}
& = &
i \, \delta^D (x - x')
\; \; , \label{app3a} \\
\partial_{\mu}^{\prime} \left\{
\sqrt{-g(x')} \; g^{\mu\nu}(x')
\partial_{\nu}^{\prime} \, {i\Delta}_{A +-}(x;x')
\right\}
& = &
0
\;\; , \label{app3b}
\end{eqnarray}
and we can trivially write:
\begin{equation}
{i\Delta}_{A}(x;x') \;
\partial_{\mu}^{\prime} \, {i\Delta}_{A}(x;x')
\; = \;
\frac12 \; \partial_{\mu}^{\prime} \, {i\Delta}^2_{A}(x;x')
\;\; , \label{app4}
\end{equation}
the partial integration gives:
\begin{eqnarray}
\left\langle \,
B^2 (t, {\vec x})
\, \right\rangle^{\mbox{\tiny 2-loop}}_{\mbox{\tiny VEV}}
& = &
{i\Delta}_A(x;x)
\biggl[ \, -\frac{\lambda}{2} \; {i\Delta}_{A}(x;x)
- \delta Z \, \biggr]
\; + \;
\label{app5} \\
& \mbox{} & \hspace{-3.8cm}
\frac{i\lambda}{4} \int d^Dx' \,
\sqrt{-g(x')} \; g^{\mu\nu}(x') \;
\partial_{\nu}^{\prime} \, {i\Delta}_{A}(x';x')
\left\{ \partial_{\mu}^{\prime} \, {i\Delta}^2_{A ++}(x;x')
- \partial_{\mu}^{\prime} \, {i\Delta}^2_{A +-}(x;x') \right\}
\nonumber 
\end{eqnarray}
Note that the only surface term is at $t'=0$ because the $++$ and $+-$
propagators cancel whenever $x^{\prime \mu}$ lies outside the
past light-cone of $x^{\mu}$. That surface term vanishes if $\delta Z$
is chosen -- as it must be -- to cancel the one loop contribution
to the self-mass squared at $t=0$:
\begin{equation}
\delta Z \; = \;
\frac{\lambda}{2} \, \frac{H^{D-2}}{(4\pi)^{\frac{D}2}} \,
\frac{\Gamma(D \!-\! 1)}{\Gamma(\frac{D}2)} \,
\pi \cot\Bigl( \frac{\pi}2 D \Bigr)
\;\; , \label{app6}
\end{equation}
Then we have:
\begin{equation}
\biggl[ -\frac{\lambda}{2} \; {i\Delta}_{A}(x';x')
- \delta Z \biggr] \; = \;
- \lambda \, \frac{H^{D-2}}{(4\pi)^{\frac{D}2}} \,
\frac{\Gamma(D \!-\! 1)}{\Gamma(\frac{D}2)} \, \ln a'
\;\; , \label{app7}
\end{equation}
which vanishes at $t'=0$.

The first term in (\ref{app5}) contributes a double infrared logarithm:
\begin{equation}
{i\Delta}_A(x;x)
\biggl[ \, -\frac{\lambda}{2} \; {i\Delta}_{A}(x;x)
- \delta Z \, \biggr]
\, = \,
-\frac{\lambda H^4}{2^5 \pi^4} \, \ln^2 a \, + \,
O(\ln a) 
\;\; . \label{app8}
\end{equation}
To see that the second term in (\ref{app5}) also contributes a double
infrared logarithm first note that the only the temporal derivative
survives:
\begin{equation}
\partial_{\nu}^{\prime} \, {i\Delta}_{A}(x';x') \, = \,
\frac{H^{D-2}}{(4\pi)^{\frac{D}2}} \, 
\frac{\Gamma(D \!-\! 1)}{\Gamma(\frac{D}2)} \, 
H a' \, \delta^0_{\nu} 
\;\; . \label{app9}
\end{equation}
Hence we can partially integrate on $\eta'$:
\begin{eqnarray}
- \frac{i\lambda}{4} \, 
\frac{H^{D-2}}{(4\pi)^{\frac{D}2}} \, 
\frac{\Gamma(D \!-\! 1)}{\Gamma(\frac{D}2)} \, 
H \int d^Dx' \, a^{\prime D-1} \; 
\frac{\partial}{\partial \eta'} \left\{ \, {i\Delta}^2_{A ++}(x;x')
- {i\Delta}^2_{A +-}(x;x') \right\} 
\nonumber \\
= \; 
\left( surface \; term \; at \; \eta' = -\frac{1}{H} \right)
\; + \hspace{3cm} \label{app10} \\
\frac{i\lambda}{4} \,
\frac{H^{D-2}}{(4\pi)^{\frac{D}2}} \, 
\frac{\Gamma(D)}{\Gamma(\frac{D}2)} \, 
H^2 \int d^Dx' \, a^{\prime D} \; 
\left\{ \, {i\Delta}^2_{A ++}(x;x') 
- {i\Delta}^2_{A +-}(x;x') \right\} 
\;\; . \nonumber
\end{eqnarray}
The surface term contributes no infrared logarithms. In another context
we have previously evaluated the leading infrared logarithm contribution 
from an integral of any power of the propagator \cite{nctrpw5}. For the 
case involving the squares one finds:
\begin{eqnarray}
\frac{i\lambda}{4} \,
\frac{H^{D-2}}{(4\pi)^{\frac{D}2}} \, 
\frac{\Gamma(D)}{\Gamma(\frac{D}2)} \, 
H^2 \int d^Dx' \, a^{\prime D} \; 
\left\{ \, {i\Delta}^2_{A ++}(x;x') 
- {i\Delta}^2_{A +-}(x;x') \right\}
\nonumber \\
= \; \frac{\lambda H^4}{2^6 \pi^4} \, \ln^2 a \, + \, O(\ln a) 
\;\; . \label{app11}
\end{eqnarray}
Consequently, the leading infrared behavior of (\ref{app5}) in $D=4$ is:
\begin{equation}
\left\langle \,
B^2 (t, {\vec x})
\, \right\rangle^{\mbox{\tiny 2-loop}}_{\mbox{\tiny VEV}}
\; = \;
\frac{\lambda H^4}{2^6 \pi^4}
\Big[ -\ln^2 a \, + \, O(\ln a) \, \Big]
\;\; , \label{app12}
\end{equation}
which is precisely the term appearing in expression (\ref{pertBd2pt}).

\end{document}